\documentclass[structabstract, a4paper]{aa}  
\usepackage{graphicx}
\usepackage{txfonts}
\usepackage[utf8]{inputenc}
\usepackage{dcolumn}
\usepackage{supertabular}
\usepackage{rotating}
\usepackage{longtable}
\usepackage{longtable, lscape}
\usepackage{lscape}
\usepackage{float}
\usepackage{subfig}
\usepackage{url}
\usepackage{color}
\usepackage{natbib}
\usepackage[dvipsnames]{xcolor}
\usepackage{ulem}
\bibpunct{(}{)}{;}{a}{}{,} 

\newcommand{\methanol}{\mbox{CH$_3$OH}}
\newcommand{\water}{\mbox{H$_2$O}}
\newcommand{\kms}{\mbox{km\,s$^{-1}$}}
\newcommand{\hco}{\mbox{H$^{13}$CO$^{+}$}(1$-$0)}

\newcolumntype{d}[1]{D{.}{\cdot}{#1}}
\newcolumntype{.}{D{.}{.}{-1}}

\newcommand{\hii}{H{\sc ii}}
\newcommand{\hiis}{H{\sc ii}s}
\newcommand{\uchii}{UC\,H{\sc ii}}
\newcommand{\uchiis}{UC\,H{\sc ii}s}
\newcommand{\hchii}{HC\,H{\sc ii}}
\newcommand{\hchiis}{HC\,H{\sc ii}s}

\newcommand{\nhthree}{NH$_3$}

\newcommand{\KS}{Kolmogorov-Smirnov}
\newcommand{\mum}{$\mu$m}

\begin{document}

\title{ATLASGAL-selected massive clumps in the inner Galaxy\\ 
\vskip 0.5cm
\Large IV. Millimeter hydrogen recombination lines from associated HII regions}
\authorrunning{Kim et al.}
\titlerunning{Millimeter Hydrogen Recombination Lines}

\author{W.-J.\,Kim\inst{1}\thanks{Member of the International Max Planck Research School (IMPRS) for Astronomy and Astrophysics at the Universities of Bonn and Cologne.}
                \and
                F.\,Wyrowski\inst{1} 
        \and
                J.\,S.\,Urquhart\inst{1,2}
         \and 
        K.\,M.\,Menten\inst{1}
        \and 
        T.\,Csengeri\inst{1}
}

\institute{ Max-Planck-Institut f\"ur Radioastronomie, Auf dem H\"ugel 69, 53121 Bonn, Germany\\ \email{wjkim@mpifr-bonn.mpg.de} \and  School of Physical Sciences, University of Kent, 
              Ingram Building, Canterbury, Kent CT2\,7NH, UK}

\date{ drafted \today\ / Received / Accepted }

\abstract
{}{ Observations of millimeter wavelength radio recombination lines (mm-RRLs) are used to search for \hii\ regions in an unbiased way that is complementary to many of the more traditional methods previously used (e.g., radio continuum, far-infrared colors, maser emission). The mm-RRLs can be used  to derive physical properties of \hii\ regions and to provide velocity information of ionized gas.}
{We  carried out targeted mm-RRL observations (39\,$\leq$ principal quantum number ($n$) $\leq$\,65 and $\Delta n$ = 1, 2, 3, and 4, named H$n\alpha$, H$n\beta$, H$n\gamma$, and H$n\delta$) using the IRAM 30m and Mopra 22m telescopes. In total, we observed 976 compact dust clumps selected from a catalog of $\sim$10,000 sources identified by the APEX Telescope Large Area Survey of the Galaxy (ATLASGAL). The sample was selected to ensure a representative mix of star-forming and quiescent clumps such that a variety of different evolutionary stages is represented. Approximately half of the clumps are mid-infrared quiet while the other half are mid-infrared bright.} 
{We  detected H$n\alpha$ mm-RRL emission toward 178 clumps; H$n\beta$, H$n\gamma$, and H$n\delta$ were also detected toward 65, 23, and 22 clumps, respectively. This is the largest sample of mm-RRLs detections published to date. Comparing the positions of these clumps with radio continuum surveys we identified compact radio counterparts for 134 clumps, confirming their association with known \hii\ regions. The nature of the other 44 detections is unclear, but 8 detections are thought to be potentially new \hii\ regions while the mm-RRL emission from the others may be due to contamination from nearby evolved \hii\ regions. Broad linewidths are seen toward nine clumps (linewidth \,$>$\,40\,\kms) revealing significant turbulent motions within the ionized gas; in the past, such wide linewidths were found toward very compact and dense \hii\ regions. We find that the systemic velocity of the associated dense molecular gas, traced by  \hco, is consistent with the mm-RRL velocities and confirms them as embedded \hii\ regions. We also find that the linewidth of the \hco\ emission is significantly wider than those without mm-RRL detection, indicating a physical connection between the embedded \hii\ region and their natal environments. We also find a correlation  between the integrated fluxes of the mm-RRLs and the 6\,cm continuum flux densities of their radio counterparts (the correlation
coefficient, $\rho$, is 0.70). By calculating the electron densities we find that  the mm-RRL emission is associated with \hii\ regions with $n_{e} < $10$^{5}$\,cm$^{-3}$ and \hii\ region diameter $>$ 0.03\,pc.}
{We  detected mm-RRLs toward 178 clumps and  identified eight new \hii\ region candidates. The broad mm-RRL from nine clumps may indicate that they arise in very young hyper-compact \hii\ regions. The mm-RRLs  trace the radio continuum sources detected by high-resolution observations and their line parameters show associations with the embedded radio sources and their parental molecular clumps.}

\keywords{surveys $-$ stars: massive $-$ stars: formation $-$ \hii\ region: ISM}
\maketitle

\section{Introduction} \label{sec:introduction} 

High-mass star formation plays a critical part in the evolution of galaxies and an important role in the chemical enrichment of the interstellar medium (ISM) in the Milky Way \citep{zinnecker2007}. High-mass star formation starts in massive cold and dense gravitationally bound clumps that are still infrared-quiet and subsequently collapse and fragment into cores. The cores contain protostars called massive young stellar objects (MYSOs) that gain their mass with high accretion rates \citep{urquhart2014_atlas}. Finally, the formation of an \hii\ region that is first hypercompact and  then ultracompact signifies the arrival of the embedded MYSO on to the main sequence and it is therefore a key stage in the evolution of massive stars. The massive stars reach the main sequence while still deeply embedded in their natal cloud and continue to accrete material even after the \hii\ region has begun to form \citep{zinnecker2007, churchwell1990, churchwell2002, kurtz2000, kurtz2005a}. Investigating \hii\ regions in their earliest stages while they are still embedded in their parental dense molecular clouds allows us to determine at what evolutionary stage high-mass stars stop  accumulating their mass $M\,>\,10\,$M$_{\odot}$ and how the accretion and outflow processes evolve after the protostar arrives on the main sequence \citep{keto2008b,churchwell2010}.

The embedded \hii\ regions are surrounded by dust cocoons, but are bright from mid- and far-infrared (mid- and far-IR) to radio wavelength since the dust absorbs ultraviolet radiation from the \hii\ regions and re-emits it at infrared  wavelengths \citep{wood1989a}. Far-IR colors, interstellar masers, and molecular lines have all been employed to identify \hii\ region candidates, and radio continuum surveys have since confirmed many of them and derived their physical parameters \citep{forster1989,wood1989a,kurtz1994,hofner1996,shepherd1996,urquhart2009,walsh2014}. A class of the densest and most compact, optically thick ionized nebula known as hyper-compact \hii\ regions (\hchiis), have been revealed by observations conducted at high radio frequencies (e.g., 43 or 50\,GHz;  \citealt{carral1997,kurtz2005b}).

These surveys have been very successful in identifying a large number of young embedded compact and ultra-compact \hii\ regions (\uchiis; \citealt{lumsden2013});  to date
$\sim$600 have been cataloged. However, they tend to target \hii\ regions with specific properties or evolutionary stages and may not provide a complete picture of the full evolutionary sequence. For example, the properties of the earliest hyper-compact phase are based on a few tens of sources;  these sources are likely to be some of the most extreme and are therefore  probably not representative of the general population of these objects.

Dust continuum emission in the submillimeter wavelength range (submm) can directly trace high column density regions in which high-mass stars form. Moreover, the cold dust absorbs IR emission from objects at different evolutionary stages and then emits thermal emission through the submm. The APEX Telescope Large Area Survey of the Galaxy (ATLASGAL;  \citealt{schuller2009}) is a dust continuum survey at 870\,$\mu$m that provides an unbiased view of the dense gas located throughout the mid-plane of the inner Galaxy ($-$60$^{\circ}$ $\leq \ell \leq$ $+$60$^{\circ}$  and |$b$| $\leq$ 1.5$^{\circ}$;  \citealt{schuller2009}). ATLASGAL has identified $\sim$10,000 dense clumps (Compact Source Catalogue (CSC;  \citealt{contreras2013,urquhart2014_atlas_cata}), GaussClump Source Catalogue (GCSC; \citealt{csengeri2014})), many of which are in the pre-stellar, protostellar, and \hii\ region stages. This sample is therefore ideal for the study of the whole embedded evolutionary sequence of massive stars.

We selected dust clumps based on flux limits at 870\,\mum\ continuum emission for follow-up molecular line surveys in the 3 mm atmospheric window using the IRAM 30m and Mopra 22m telescopes. The clumps were originally selected as the brightest 870\,\mum\ clumps being mid-IR dark and bright, respectively, and later divided into three photometric categories (i.e., 22\,\mum\ dark, bright, and \hii\ regions; \citealt{csengeri2016_sio}). The clump photometric categories were determined from their association with mid-IR emission from the all-sky Wide-field Infrared Survey Explorer (WISE; \citealt{Wright2010}) and the Galactic Legacy Infrared Mid-Plane Survey Extraordinaire (GLIMPSE ; \citealt{benjamin2003_ori}). 
Furthermore, the flux limited selection of clumps with different infrared properties ensures that for the different evolutionary stages the most massive clumps were observed \citep{giannetti2014,csengeri2016_sio}.  

Although the mid-IR surveys are useful for distinguishing between IR bright (protostellar) and IR dark (starless) sources, they cannot identify \hii\ regions and therefore additional information is required. The hot ionized regions emit strongly at radio wavelengths and can be identified using radio continuum observations. The CORNISH survey \citep{hoare2012,purcell2013}
has been used to identify $\sim$200 \hii\ regions associated with ATLASGAL clumps (\citealt{urquhart2013b}) and the targeted radio continuum observations conducted as part of the RMS survey (\citealt{urquhart2007,urquhart2009}) have identified several hundreds more; a catalog of these matches and a summary of their properties is given in \citet{urquhart2014_atlas}. All of these \hii\ regions have been identified from 5-9\,GHz radio continuum observations. 
These observations are sensitive to compact and ultra-compact (UC) \hii\ regions that are generally optically thin at these frequencies, while they are not sensitive to the more embedded, optically thick, hyper-compact (HC) \hii\ regions.
However, these regions are optically thin at millimeter wavelengths and so mm-RRLs offer an opportunity to identify a younger generation of embedded \hii\ regions that may have been missed from radio continuum surveys.

Most of the previous radio recombination line surveys have been made with single-dish telescopes at centimeter (cm) wavelengths. The cm-RRLs have principal quantum numbers ($n$) $\geq$ 85 and were mostly observed with angular resolutions of a few arc-minutes (e.g., \citealt{lockman1989,caswell1987,anderson2009a,anderson2014,alves2015}). There are a number of higher resolution studies of individual sources \citep{gaume1995,depree2004,sewilo2004b,sewilo2008,keto2008a}. These studies have provided some insight into the properties of \hii\ regions, and have revealed that they   have significant broad  linewidths that decrease as the \hii\ region expands. The broadest linewidths are found toward the youngest \hii\ regions  (\hchiis\,$>$\,40\,\kms), while the slightly more evolved \uchiis\ have typical linewidths between 30\,$-$\,40\,\kms. 

The Mopra and IRAM molecular line surveys cover the frequencies of a number of mm-RRL transition. Such RRLs have been observed in a few targeted studies (e.g., \citealt{jaffe1999,churchwell2010}). These tend to focus on individual or small samples of sources.
In this paper we use the mm-RRLs observed in our line surveys to conduct a search for deeply embedded \hii\ regions in an untargeted and relatively unbiased way given that the clumps were selected based on their submillimeter and mid-infrared properties alone and not specifically with \hii\ regions in mind. These observation, therefore, have the  potential to identify new \hii\ regions and provide information on the kinematics of the ionized gas and their interaction with the molecular gas of their natal clump; this  in turn will improve our understanding of how the \hii\  regions impact the physical structure and dynamics of their local environment.

Since mm-RRL sources have so far not been studied in a  systematic way, the low sample statistics makes it difficult to compare their properties with those determined from cm-RRL studies. In fact, cm- and mm-RRLs are likely to be probing different evolutionary stages and/or different physical conditions. For example, the linewidths of cm-RRLs are affected by pressure broadening caused by high electron density; this leads to a significant broadening of the line profile \citep{gordon2002,sewilo2008}. 
However, the impact of pressure broadening for mm-RRLs is negligible and thus the mm-RRLs  linewidths are dominated by thermal and turbulent motions within the gas \citep{gordon2002}. 
Therefore, mm-RRLs provide a useful probe for the  study of the intrinsic motions and physical properties of compact \hii\ regions. Furthermore, observations of different order transitions (e.g., $\Delta n$ = 1, 2, 3, $\ldots$, denoted H$n\alpha$, H$n\beta$, H$n\gamma$, and so on) allow us to determine whether the mm-RRL emission departs from non-LTE or LTE conditions since the degree of such departures may change with quantum number and order \citep{thum1995}.

In this paper we present a new mm-RRL survey to find embedded \hii\ regions in ATLASGAL clumps and derive the properties of the \hii\ regions detected. This mm-RRL survey provides the largest sample of mm-RRL detections and an unbiased way to search for \hii\ regions. The structure of the paper is as follows. The observations and data reduction is described in Sect.\,\ref{sec:observation}. The general results of detected mm-RRLs and association with molecular clumps, radio continuum, and mid-IR counterparts are presented in Sect.\,\ref{sec:result}. The relationships with the radio continuum counterparts and calculated physical parameters are discussed in Sect.\,\ref{sec:analysis}. Finally, we summarize our main results in Sect.\,\ref{sec:summary}.

\begin{table*}
\centering
\tiny
\caption{\label{tb:obs_log} Summary of the observational setup of the telescopes, coverage, and the number of sources observed.}
\begin{tabular}{ c c c c c c c c .}
\hline \hline
Frequency  &  Telescope & Date  & $\ell$ & Number of & rms & $\Delta v$ resolution & Beam size &  \multicolumn{1}{c}{K to Jy} \\      
(GHz)   &    &    (mm/yy)    &  (deg) & sources   &  (Jy)   &  (\kms) &($''$)&  \multicolumn{1}{c}{(Jy\,K$^{-1}$)} \\
\hline
85.6\,$-$\,88.4&Mopra 22m&05/08, 08/09 \& 09/09& $-$ 60$^{\circ}\leq$\,$\ell$\,$\leq$\,0$^{\circ}$& 566 & 0.20& 3.5$\sim$3.8&$\sim$36&22.0\\
85.6\,$-$\,110.6&IRAM 30m&04/11 \& 10/12 &0$^{\circ}\leq$\,$\ell$\,$\leq$\,$+$60$^{\circ}$&    410 & 0.05& 3.7$\sim$4.1 &  $\sim$29&5.9\\
\hline
\end{tabular}
\end{table*}

\section{Observations and data reduction}\label{sec:observation}

\subsection{Source selection}
The selection of the sources for the Mopra and IRAM surveys was based on their 870\,\mum\ continuum peak flux. We applied different flux limits to clumps with different infrared properties, to cover a range of evolutionary stages, and observed all sources above the flux limits. For the earlier Mopra survey, we used the 21\,\mum\ emission properties, 21\,\mum-bright and -dark, of the Midcourse Space Experiment (MSX; \citealt{price2001_msx}), and when Spitzer data became available, we used their  8 and 24\,\mum\ emission properties for the IRAM survey (see \citealt{giannetti2014} and \citealt{csengeri2016_sio} for details).

As the results, the 870\,\mum\ peak flux thresholds are different for the Mopra and IRAM surveys. 
Nevertheless, both surveys can be considered to cover all different evolutionary stages and differ slightly by the mass ranges they cover.
We use all the sources of both surveys as a single sample for all the data analysis, and the number of sources (976) are large enough to reduce potential biases.

\subsection{Observational setups}

We  used the Mopra 22m and IRAM 30m telescopes to observe a sample of ATLASGAL source distributed across the inner Galactic plane. The beam sizes and frequency coverage results in different angular resolutions and sensitivities for the two telescopes; they are summarized in Table\,\ref{tb:obs_log}. Since we have stacked the mm-RRLs to improve the signal-to-noise ratio (S/N)  of the observations, in the following we use  the beam size of the lowest frequency transition; this is the value given in Table\,\ref{tb:obs_log}. The number of sources observed in different longitude ranges by the Mopra 22m and the IRAM 30m telescopes (see next section) are also summarized in Table\,\ref{tb:obs_log}.

\subsubsection{Mopra 22m telescope observations}
The mm-RRL data for the southern hemisphere were obtained from targeted molecular line observations with the Mopra 22m telescope (Project IDs: M327-2008 and M327-2009; \citealt{wyrowski2008,wyrowski2009}).\footnote{The Mopra radio telescope is part of the Australia Telescope National Facility which is funded by the Australian Government for operation as a National Facility managed by CSIRO.} 
As previously mentioned, the sources were selected based on their mid-IR emission properties (mid-IR bright or mid-IR dark) to ensure that a range of evolutionary stages was probed. The 870\,\mum\ peak flux threshold was based on 
the IR properties of the clumps; for mid-IR bright clumps a value of 1.75\,Jy\,beam$^{-1}$ was used, while for the mid-IR dark clumps flux limit of 1.2\,Jy\,beam$^{-1}$ was used. In total, 566 ATLASGAL dust clumps were observed within the longitude ranges $-$60$^{\circ} \leq \ell \leq$ 0$^{\circ}$. 
The Mopra observations used the MOPS spectrometer\footnote{The University of New South Wales Digital Filter Bank used for the observations with the Mopra Telescope was provided with support from the Australian Research Council.}, which has a bandwidth coverage of 8\,GHz and was tuned to 89.3 GHz providing coverage of 85.2 -- 93.4\,GHz with
a velocity resolution of $\sim$0.9\,\kms\ and angular resolution of $\sim$\,36$\arcsec \pm 3\arcsec$ at 86\,GHz (\citealt{ladd2005}). This frequency range includes three mm-RRL transitions, H41$\alpha$, H42$\alpha$, and H52$\beta$. We have scaled the measured antenna temperatures, $T_{\rm{a}}^{*}$, to the main-beam brightness temperatures, $T_{\rm MB}$, using the Mopra main-beam efficiency ($\eta_{\rm MB}$\,=\,0.49; \citealt{ladd2005}).

\subsubsection{IRAM 30m telescope observations}
For the northern hemisphere the mm-RRL data were obtained from targeted molecular line observations with the IRAM 30m telescope (Project IDs: 181-10 and 037-12) (see \citealt{csengeri2016_sio} for more details).\footnote{IRAM is supported by INSU/CNRS (France), MPG (Germany), and IGN (Spain).} 
The 870\,\mum\ peak flux thresholds of mid-bright and mid-IR dark clumps for the IRAM 30m observations were 0.6\,Jy\,beam$^{-1}$ and 0.3\,Jy\,beam$^{-1}$, respectively. 
In total, 410 ATLASGAL dust clumps were observed within the longitude ranges 0$^{\circ} \leq$ $\ell$ $\leq$ $+$60$^{\circ}$. 
The IRAM observations used the EMIR receiver, which covers almost the whole 3 mm atmospheric window with a velocity resolution of $\sim$0.5\,\kms. These observations were therefore able to observe twenty mm-RRL transitions including the three observed with Mopra. The observed mm-RRL transitions, rest frequencies, and absorption oscillator strengths are listed in the first three columns of Table\,\ref{tb:info_rrls}. We  used the forward efficiency ($\eta_{l}$\,=\,0.95) and the main-beam efficiency ($\eta_{\rm MB}$\,=\,0.81) to convert the  $T_{\rm{a}}^{*}$ to the $T_{\rm MB}$ for the IRAM 30m observations\footnote{http://www.iram.es/IRAMES/mainWiki/Iram30mEfficiencies}.

\begin{figure}
\begin{center}
\includegraphics[width=0.50\textwidth, trim= 0cm 0.6cm 0.5cm 0cm, clip]{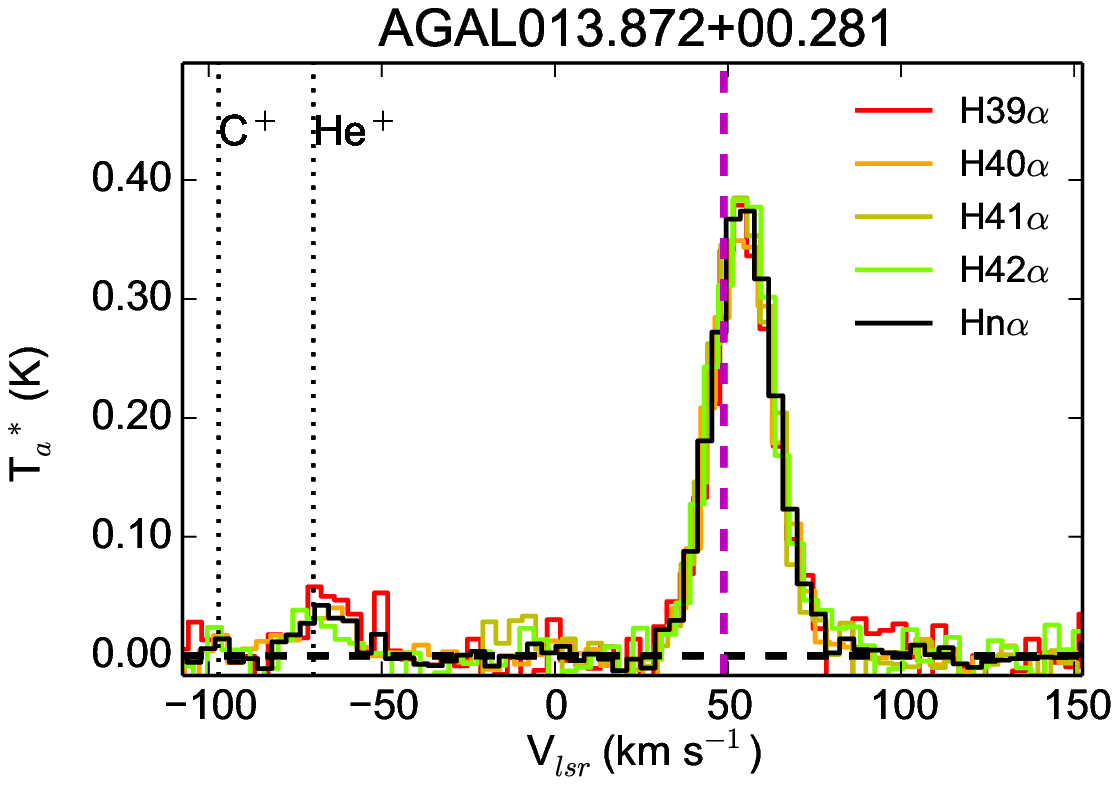}
\includegraphics[width=0.50\textwidth, trim= 0cm 0.6cm 0.5cm 0cm, clip]{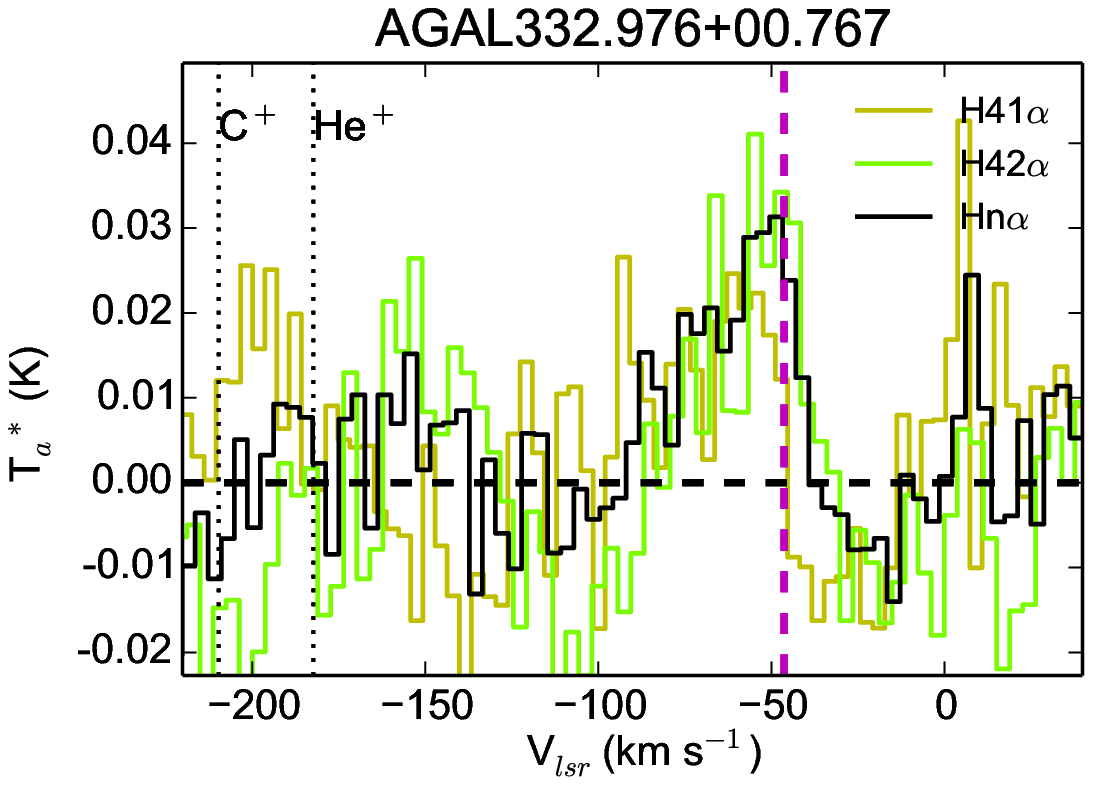}
\caption{\label{fig:stacked} For two sources in our sample, each color represents an observed RRL H$n\alpha$ transition and the black line shows the stacked spectrum. The purple dashed lines indicate the systemic local standard of rest (lsr) velocity of the clump determined from the \hco\ transition and the two vertical black dotted lines indicate expected positions of helium and carbon RRLs.}
\end{center}
\end{figure}
\begin{table*}
\small
\centering
\caption{\label{tb:info_rrls}Frequency, absorption oscillator strength, and detection rates of observed hydrogen millimeter recombination lines. The H$n\beta$, H$n\gamma$, and H$n\delta$ mm-RRLs are only detected toward clumps where the peak intensities of H$n\alpha$ mm-RRLs are brighter than 0.24, 0.62, and 0.54\,Jy, respectively.}
\begin{tabular}{ c.cccc}
\hline \hline
mm-RRL & \multicolumn{1}{c}{Rest frequency} & \multicolumn{1}{c}{Absorption} & Number of & Number of  &  Detection Rate\tablefootmark{b} \\
Transition & \multicolumn{1}{c}{(MHz)} & \multicolumn{1}{c}{oscillator strength$^{*}$} & Observed sources & Detected sources\tablefootmark{a}& \\
\hline
H$n\alpha$& $--$& $-$ &976 &  178 & 18\,$\pm$\,1  \\     
H39$\alpha$& 106737.4& 7.725489&385 &  58 & 15\,$\pm$\,2  \\ 
H40$\alpha$& 99023.0& 7.916287&366 &  57 & 16\,$\pm$\,2 \\
~~H41$\alpha$\tablefootmark{$\dagger$}& 92034.4& 8.107084&972 &  ~~146\tablefootmark{c} & 15\,$\pm$\,1   \\ 
~~H42$\alpha$\tablefootmark{$\dagger$}& 85688.4& 8.297880&972 &  136 &  14\,$\pm$\,1  \\ 
\hline
H$n\beta$ &  $--$ & $-$ &976 &  65 &  7\,$\pm$\,1 \\
H48$\beta$ &  111885.1& 1.342709&366 &  20 & 5\,$\pm$\,1 \\ 
H49$\beta$ &  105301.9& 1.369046&385 &  26 & 7\,$\pm$\,1  \\ 
H50$\beta$ &  99225.2& 1.395383&366 &  28 &  8\,$\pm$\,1 \\ 
H51$\beta$ &  93607.3& 1.421720&366 &  28 & 8\,$\pm$\,1 \\ 
~~H52$\beta$\tablefootmark{$\dagger$} &  88405.7& 1.448056&971 & 47 &  5\,$\pm$\,1 \\ 
\hline 
H$n\gamma$& $--$ & $-$ &389 &  23 &  6\,$\pm$\,1\\
H55$\gamma$&  109536.0& 0.482179&366 & 17  &  5\,$\pm$\,1 \\ 
H56$\gamma$&  103914.8& 0.490287&385 & 17  &  4\,$\pm$\,1 \\ 
H57$\gamma$&  98671.9& 0.498395&366 & 18   & 5\,$\pm$\,1 \\ 
H58$\gamma$&  93775.9& 0.506503&366 & 19  &  5\,$\pm$\,1 \\ 
H60$\gamma$&  84914.4& 0.522718&385 & 15  &  4\,$\pm$\,1\\ 
\hline
H$n\delta$& $--$ & $-$ &389 &  22  &  6\,$\pm$\,1 \\
H60$\delta$&  110600.7& 0.230388&366 & 9  &  2\,$\pm$\,1 \\ 
H61$\delta$&  105410.2& 0.233881&385 & 9  &  2\,$\pm$\,1 \\ 
H62$\delta$&  100539.6& 0.237374&385 & 11  & 3\,$\pm$\,1  \\ 
H63$\delta$&  95964.6& 0.240867&366 & 10  & 3\,$\pm$\,1 \\ 
H64$\delta$&  91663.1& 0.244359&384 & 11 &  3\,$\pm$\,1 \\
H65$\delta$&  87615.0& 0.247852&383 &  7 &  2\,$\pm$\,1 \\ 
\hline
\end{tabular}
\tablebib{(*)~\citet{goldwire1968, menzel1969}.}
\tablefoot{
{Column 1 is mm-RRLs increasing transition, in Cols. 2 and 3 we give the numbers of observed and detected sources, respectively. The last column shows the detection rates. Line parameters and rms levels of the stacked and individual transitions are available in electronic form at the CDS.}
\tablefoottext{$\dagger$}{Mopra 22m covered only these three transitions.}
\tablefoottext{a}{A detection requires a stacked signal  above 3\,$\sigma$.}
\tablefoottext{b}{The statistical errors of detection rates are obtained assuming binomial statistics.}
\tablefoottext{c}{Some H41$\alpha$ spectra are located at an edge or beyond observed band width.}}
\end{table*}

\subsection{Data reduction}
The data reduction was performed using the CLASS program of the GILDAS package\footnote{https://www.iram.fr/IRAMFR/GILDAS/doc/html/class-html/class.html}. Since the RRLs with the adjacent principal quantum numbers ($n$) and the same $\Delta n$ have similar energy levels and absorption oscillator strengths, resulting in similar intensities, it is possible to stack the spectra to improve the S/N of the emission. This technique has been successfully used in cm-RRLs surveys (e.g., \citealt{alves2010,anderson2011}). We show two examples of the stacked spectra in Fig.\,\ref{fig:stacked}; the upper panel shows the four H$n\alpha$ transitions ($n$ = 39, 40, 41, and 42) of AGAL013.872$+$00.281, while the lower panel shows the H$n\alpha$ transitions ($n$ = 41 and 42) emission detect toward AGAL332.976+00.767. In both panels we show the profile of the combined stacked spectra (black line). The emission seen toward AGAL013.872$+$00.281 is clearly detected; however,  the emission toward AGAL332.976+00.767  is ambiguous in the individual transitions but is detected at a significant level in the stacked line ($>3\sigma$). 
After stacking all transitions, a polynomial baseline of order 1 to 3 fitted to a 200\,\kms\ wide line-free velocity range was subtracted from the spectra. 

In addition to the mm-RRLs, we  also used the high number density and high column density tracing \hco\ ($\nu$ = 86754.330\,MHz) and N$_2$H$^+$ (1$-$0) ($\nu$ = 93173.772\,MHz) transitions to determine the systemic radial velocities of the associated dense molecular clumps. The velocities of the peak emission in the \hco\ and N$_2$H$^+$ (1$-$0) line spectra were determined using CLASS and are indicated in Fig.\,\ref{fig:stacked} by the vertical purple dashed line. 

\section{Results}\label{sec:result}

\subsection{Detection rates}\label{sec:detection}
\begin{figure}[!htbp]
\begin{center}
\includegraphics[width=0.45\textwidth, trim= 0.5cm 2.5cm 0.5cm 1.8cm, clip]{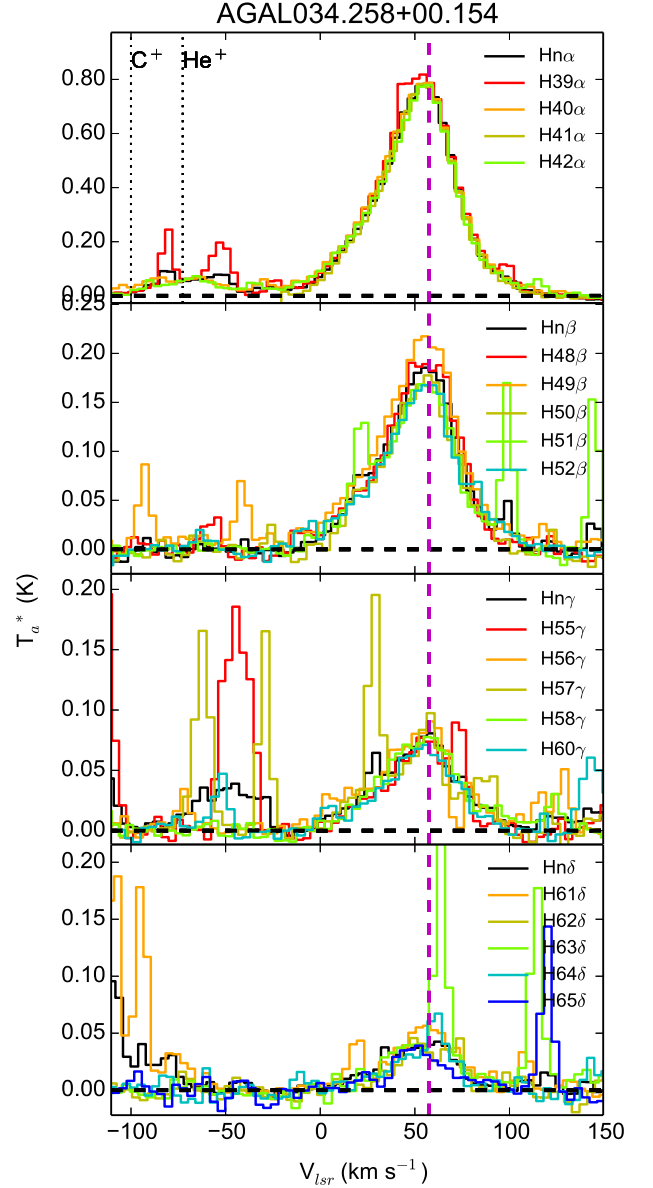}
\caption{\label{fig:2_spec} One of the mm-RRL spectra detected toward the ATLASGAL sources. Each color represents a different transition. The black lines are stacked mm-RRLs and the purple dashed lines indicate the velocity of the \hco\ emission line. The vertical black dotted lines indicate positions of expected helium and carbon RRLs. Other emission lines are an unidentified molecular emission. }
\end{center}
\end{figure}

Fitting Gaussian profiles to spectra of the stacked and individual transitions provided line parameters and rms levels. We used a peak intensity greater than 3\,$\sigma$  as a threshold for detections.  
In total, we  detected H$n\alpha$ emission toward 178 clumps and H$n\beta$, H$n\gamma$, and H$n\delta$ RRLs  toward 65, 23, and 22 clumps, which represent 18\%, 7\%, 6\%, and 6\% of the full sample, respectively. 
Figure\,\ref{fig:2_spec} shows one object in the sample where mm-RRL emission is  detected in H$n\alpha$, H$n\beta$, H$n\gamma$, and H$n\delta$ transitions.
The numbers of observed and detected sources, and detection rates of stacked lines (H$n\alpha$, H$n\beta$, H$n\gamma$, and H$n\delta$) and individual transition lines are summarized in Table\,\ref{tb:info_rrls}.
Variations in the numbers of observed sources in the individual transition lines occur because the IRAM 30m observation did not cover all transitions toward all clumps targeted.
Clumps with detected mm-RRL are listed in Table\,\ref{tb:mmrrl_list} which also provides flags to indicate their association with mid-IR and radio continuum emission (these associations are discussed in more detail in a later section).

In general, the relative intensity of RRL emission changes with $\Delta n$ of the RRL. 
The intensities of small $\Delta n$ RRLs (i.e., H$n\alpha$ transition) are higher than the intensities of larger $\Delta n$ RRLs (i.e., H$n\beta$, H$n\gamma$, and H$n\delta$ transitions) \citep{towle1996}. This explains the lower detection rates of $\Delta n >$ 1 RRLs for a given sensitivity.  Furthermore, the H$n\alpha$ transition can be stimulated more easily by maser emission than higher-order transitions under non-LTE conditions, which can lead to an increase in their intensity and detection rate. In Sect.\,\ref{sec:excitation}, we  discuss whether the H$n\alpha$ mm-RRLs depart from LTE.

\setlength{\tabcolsep}{4pt}

\begin{table*}
\tiny
\centering
\caption{\label{tb:mmrrl_list} Clumps detected with mm-RRLs.} 
\begin{tabular}{c c c c . c c c c c c c}
\hline \hline
ID & ATLASGAL &RA& Dec. & \multicolumn{1}{c}{Dist} & Log($M_{\rm clump}$) & Log($L_{\rm bol}$) & \multicolumn{1}{c}{H$n\alpha$ v$_{\rm lsr}$}& \multicolumn{1}{c}{ Systemic v$_{\rm lsr}$} & WISE & Radio & Comments \\
No. & clump name& $\alpha$(J2000) & $\delta$(J2000) & \multicolumn{1}{c}{(kpc)} & (M$_\odot$) & (L$_\odot$) & \multicolumn{1}{c}{(\kms)} & \multicolumn{1}{c}{(\kms)}&mid-IR & emission& \\
\hline
1&AGAL008.671$-$00.356 & 18:06:19.0 & $-$21:37:28 &$--$&$-$&$-$& $+$44 &$+$35& IR-bright & Y&EGO \\
2&AGAL010.151$-$00.344 & 18:09:21.2 & $-$20:19:28 &1.6&2.11&$-$& $+$20 &$+$9& Complex & Y& \\
3&AGAL010.168$-$00.362 & 18:09:26.7 & $-$20:19:03 &3.6&3.01&$-$& $+$10 &$+$14&Complex &  Y&W31 \\
4&AGAL010.323$-$00.161 & 18:09:01.4 & $-$20:05:12 &3.5&3.60&4.62& $+$5 &$+$12& IR-bright        & Y& \\
5&AGAL010.472$+$00.027 & 18:08:37.9 & $-$19:51:48 &8.6&4.55&5.34& $+$60 &$+$67& IR-bright        & Y& \\
6&AGAL010.624$-$00.384 & 18:10:28.6 & $-$19:55:46 &4.9&4.20&5.61& $+$0 &$-$3& IR-bright & Y&W31 \\
7&AGAL010.957$+$00.022 & 18:09:39.2 & $-$19:26:28 &13.7&4.15&5.10& $+$18 &$+$21& IR-bright &     Y&  \\
8&AGAL011.034$+$00.061 & 18:09:39.7 & $-$19:21:20 &14.4&3.84&5.02& $+$9 &$+$15& IR-bright        &  Y&\\
9&AGAL011.936$-$00.616 & 18:14:00.8 & $-$18:53:24 &4.0&3.61&4.97& $+$42 &$+$38& IR-bright        &  Y&\\
10&AGAL012.208$-$00.102 & 18:12:39.6 & $-$18:24:14 &13.6&4.58&$-$& $+$28 &$+$24& IR-bright & Y&  \\
11&AGAL012.804$-$00.199 & 18:14:13.5 & $-$17:55:32 &2.4&4.03&5.11& $+$36 &$+$36& IR-bright & Y& W33 \\
12&AGAL013.209$-$00.144 & 18:14:49.3 & $-$17:32:46 &4.6&3.55&$-$& $+$50 &$+$52& IR-bright & Y&  \\
13&AGAL013.384$+$00.064 & 18:14:24.9 & $-$17:17:39 &1.9&2.02&$-$& $+$14 &$+$14& IR-bright & Y&  \\
14&AGAL013.872$+$00.281 & 18:14:35.6 & $-$16:45:39 &4.4&3.50&$-$& $+$52 &$+$49& IR-bright & Y&  \\
15&AGAL015.013$-$00.671 & 18:20:21.3 & $-$16:12:42 &2.0&3.84&4.13& $+$8 &$+$18& Complex& C & M17 \\
16&AGAL015.024$-$00.654 & 18:20:17.9 & $-$16:11:30 &2.0&2.73&$-$& $+$20 &$+$19& Complex & C & M17 \\
17&AGAL015.029$-$00.669 & 18:20:22.4 & $-$16:11:44 &2.0&3.68&5.17& $+$10 &$+$19& Complex & C & M17SW \\
18&AGAL015.051$-$00.642 & 18:20:18.7 & $-$16:09:43 &2.0&$-$&$-$& $+$21 &$+$18& Complex & C & M17 \\
19&AGAL015.056$-$00.624 & 18:20:15.4 & $-$16:08:59 &2.0&2.17&$-$&$+$22&$+$18& Complex & C & M17 \\
\hline
\end{tabular}
\tablefoot{Only a portion of the entire table is given here for guidance of content. The full table is available in electronic form at the CDS. Columns, from left to right, are source ID, equatorial coordinates, heliocentric distance, ATLASGAL clump mass, bolometric luminosity of all RMS sources embedded in the ATLASGAL clump, local standard of rest (lsr) velocity of the mm-RRL and the systemic velocity of the dense clump (as determined from the \hco\ or N$_2$H$^+$ (1$-$0) transitions), mid-IR property, presence of radio emission, and comments. For radio emission, the flags Y and N indicates whether or not a radio continuum source is found within a radius of $18''$.  A flag C indicates possible contamination from nearby extended radio sources that are located within a radius of $2'$.} 
\end{table*}

\setlength{\tabcolsep}{6pt}


\begin{table*}
\small
\centering
\caption{\label{tb:inten_width}Intensity and linewidth of mm-RRLs.}
\begin{tabular}{ c c c  c  c  c c }
\hline \hline
 Stacked line & No. of sources & \multicolumn{2}{c}{S$_{\rm peak}$ (Jy)} & \multicolumn{3}{c}{ linewidth (\kms)}\\
Transition&  & Mean (rms) & Median (rms) & Mean & $\sigma$ & Standard error\\
\hline
H$n\alpha$ & 178 &1.75 (0.14) & 0.91 (0.15) & 28.4 & 7.0 &0.52\\
H$n\beta$ & 65 &1.02 (0.14) & 0.70 (0.06) & 28.3 & 7.4 & 0.92\\
H$n\gamma$ & 23 &0.37 (0.04) & 0.29 (0.04) & 28.5 & 7.5 & 1.57\\
H$n\delta$ & 22 &0.25 (0.04) & 0.18 (0.04) & 27.2 & 8.4 & 1.80\\
\hline \hline
\end{tabular}
\end{table*}

\begin{figure}
\begin{center}
\includegraphics[width=0.48\textwidth]{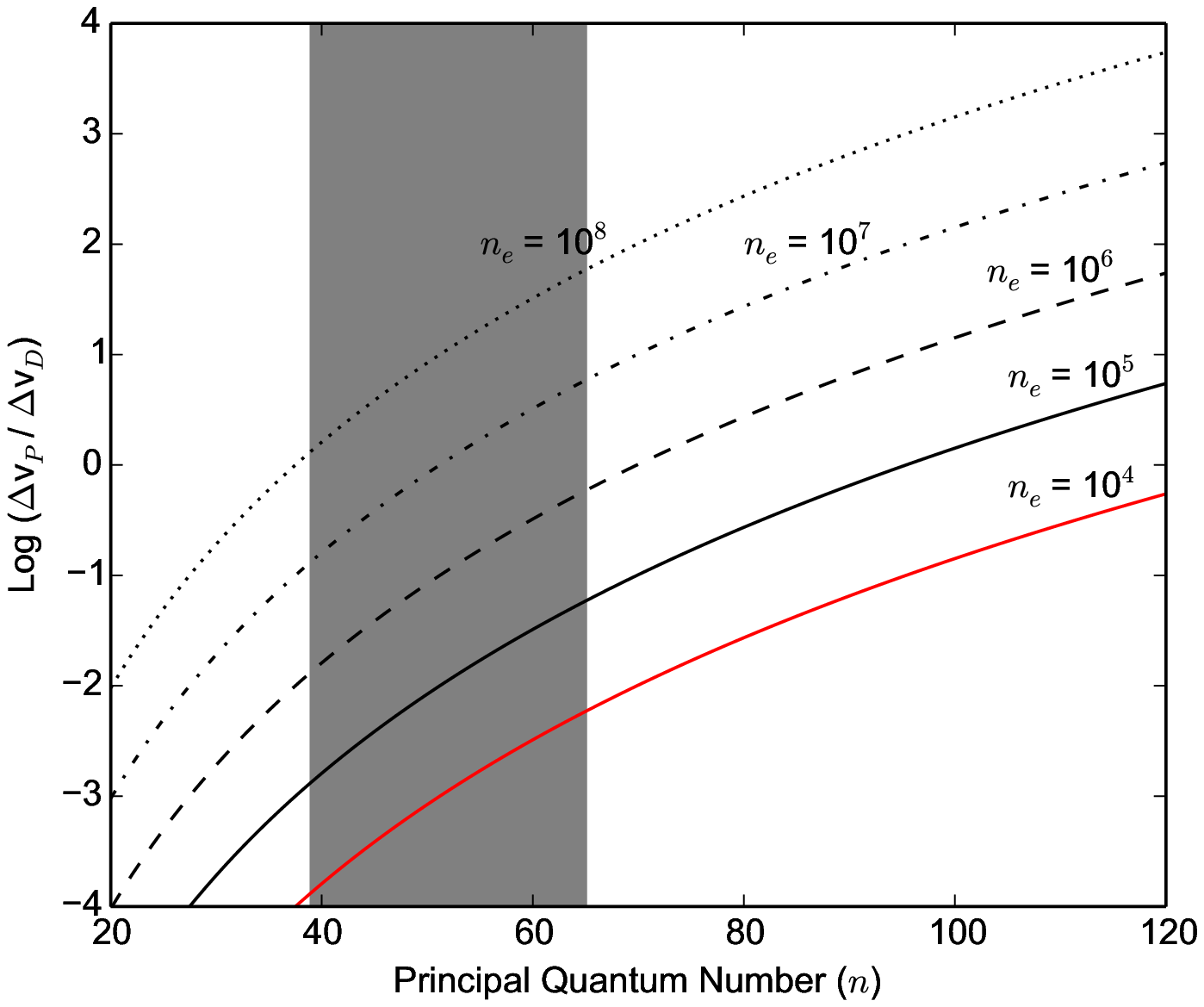}
\caption{\label{fig:n_ne} Plot of Eq.\,(4.9) from \cite{brocklehurst1972} with different electron densities ($n_{e}$). The gray area indicates observed mm-RRLs transitions (39\,$\leq n \leq$\,65) in this paper. The curves show how the relationship between the linewidth ratio and the principal quantum numbers varies as a function of the electron density. The values $\Delta$v$_{P}$ and $\Delta$v$_{D}$ indicate the linewidths produced by pressure broadening and doppler broadening effects, respectively.}
\end{center}
\end{figure}

\subsection{General properties of the mm-RRLs}\label{sec:lines}

\begin{figure}
\begin{center}
\includegraphics[width=0.48\textwidth]{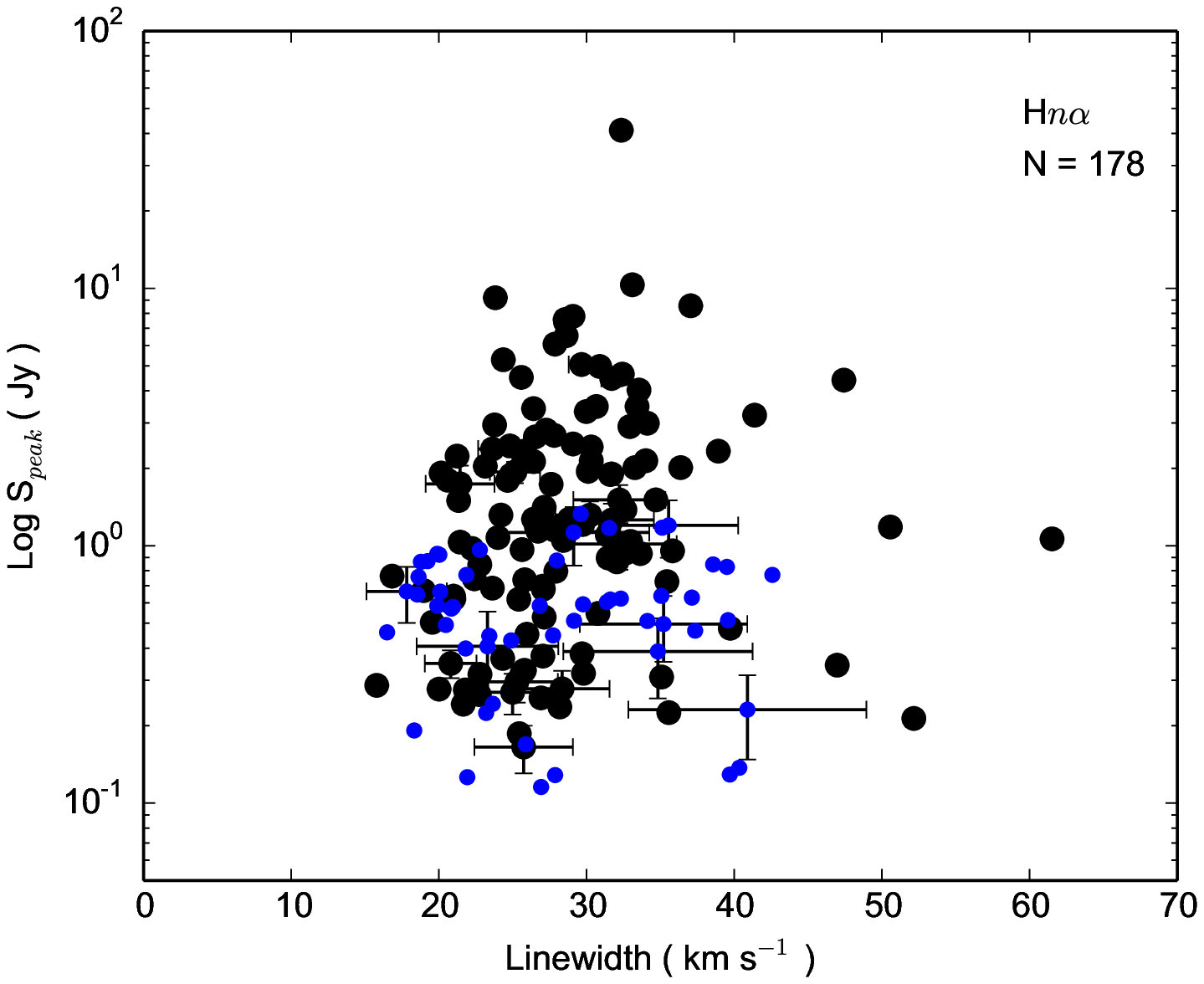}
\caption{\label{fig:inten_width} Distribution of H$n\alpha$ RRLs peak intensity as a function of linewidth. The blue dots indicate sources with mm-RRLs detected below  5\,$\sigma$. The error bars are plotted for every 7th source chosen at random to show their range. }
\end{center}
\end{figure}

In Table\,\ref{tb:inten_width} we present a summary of the mean intensities and linewidths of the various stacked mm-RRL spectra.  The mean and median peak intensities clearly decrease toward higher-order transitions while the mean linewidths for all transitions are similar (linewidth $\sim$30\,\kms).
Figure\,\ref{fig:n_ne} shows how pressure broadening affects the observed linewidths as a function of the principal quantum number ($n$) and electron density.
Even though there is a variation in the pressure broadening in the range of our observed transitions (gray filled area) for high electron densities ($\geq$ 10$^{5}$\,cm$^{-3}$), the pressure broadening of mm-RRLs from \uchiis\ with typical density of 10$^{4}$\,cm$^{-3}$ \citep{kurtz2005b} is very small and can be neglected. The linewidths of mm-RRLs for all transitions considered here are, therefore, likely to be dominated by unresolved turbulent and thermal motions of the ionized gas. 
In addition, the linewidth ratio of higher-order to H$n\alpha$ transitions shows whether there is any pressure broadening affecting the mm-RRL \citep{viner1979}: Mean linewidth ratios of H$n\beta$/H$n\alpha$, H$n\gamma$/H$n\alpha$, and H$n\delta$/H$n\alpha$ are 0.96$\pm$0.12, 0.93$\pm$0.10, and 0.89$\pm$0.15, respectively. Since the pressure broadening effect increases with principal quantum number ($n$) we  expect to find significant variations from unity if there is significant pressure broadening in the gas; however, this is not seen suggesting that this effect is negligible.
The relation of peak intensities and the linewidths of the H$n\alpha$ mm-RRLs is shown in Fig.\,\ref{fig:inten_width}.  In general, evolved \hii\ regions such as \uchii\ and compact \hii\ regions have been found to have linewidths of  $\sim$30\,\kms; however, broader RRL have been reported toward \hchiis\ (linewidth $\geq$\,40\,\kms; e.g., \citealt{sewilo2004b}). In  Fig.\,\ref{fig:inten_width}, the linewidths of the majority of mm-RRLs are narrower than 40\,\kms, which is consistent with the RRLs being associated with \uchii\ regions. However, this plot also reveals nine clumps that have broader mm-RRL linewidths ($>$\,40\,\kms), six  of which have a S/N $>$\,5 and so are considered to be reliable. Such broad linewidths are normally found toward more compact \hii\ regions making these six sources potential \hchii\ region candidates (source ID: N32, N49, N51, N62, N73, and N107 in  Table\,\ref{tb:mmrrl_list}) 
and we  investigate some of these in more detail in Sect.\,\ref{sec:hchii}.

\subsubsection{Excitation conditions}\label{sec:excitation}

Comparing peak intensity ratios of H$n\alpha$ and higher-order RRLs is useful in order to diagnose whether the mm-RRLs are emitted under local thermodynamic equilibrium (LTE) or non-LTE conditions.  Table\,\ref{tb:ratio_rrls} gives the measured mean ratios from the mm-RRLs and the expected ratios assuming LTE condition of pairs of transitions. In non-LTE conditions the measured ratio is lower than expected under LTE conditions, due to an increase in the H$n\alpha$ intensity from stimulated maser amplification than the higher-order lines, as seen in MWC349A \citep{thum1995}.

The peak intensity ratios for the observed mm-RRLs are, on average, consistent with the corresponding LTE ratios within the 1\,$\sigma$ scatter of the observed values. 
However, we note that the mm-RRLs from AGAL034.258+00.154 (ID: N49) and AGAL043.166+00.011 (ID: N55) show significantly lower peak intensity ratios (e.g., H$n\beta$/H$n\alpha$ (0.236$\pm$0.008 and 0.252$\pm$0.007), H$n\gamma$/H$n\alpha$ (0.096$\pm$0.008 and 0.102$\pm$0.007) and H$n\delta$/H$n\alpha$ (0.055$\pm$0.005 and 0.064$\pm$0.005)). 
A geometrical effect in a spherical \hii\ region with variable electron temperature and density could produce these low ratios \citep{walmsley1990}. 

Since these transitions are observed with similar spatial resolution, the geometrical effect is not important. 
The H$n\alpha$ RRLs of the two sources might be enhanced by weak maser amplification even though their ratios do not deviate from LTE to the same extent as the well-known case of MWC349A (see \citealt{thum1995,martin-pintado2002}).
Therefore, these two \hii\ regions are candidates for new RRL maser sources and observations are underway to investigate these sources in more detail. The results of these observations will be reported in a future paper.

\begin{table}
\small
\centering
\caption{\label{tb:ratio_rrls} Ratio of mm-RRL transitions. The (\#) indicates the number of sources that are detected in both transitions. These mean values are average values of individually measured ratios of sources in which both lines are detected.}
\begin{tabular}{ c  c  c }
\hline \hline
 & \multicolumn{2}{c}{Mean peak intensity ratio} \\
Pairs (\#) & Observed & Under LTE \\
\hline
H$n\beta$/H$n\alpha$ (65) & 0.31\,$\pm$\,0.07& 0.27 \\
H$n\gamma$/H$n\alpha$ (23) & 0.14\,$\pm$\,0.02 & 0.13 \\
H$n\delta$/H$n\alpha$ (22) & 0.10\,$\pm$\,0.02 & 0.07 \\
\hline 
\end{tabular}
\end{table}

\subsection{Systematic velocities of the clumps}
\label{sect:velocity_comparison}

We  used the \hco\ and N$_{2}$H$^{+}$ (1$-$0) transitions as a tracer of the systematic velocity of the dense molecular clumps \citep{beuther2007}. The \hco\ emission was detected toward 170 sources, and so the N$_{2}$H$^{+}$ (1$-$0) transition was used to determined the velocities for the other eight sources.  Both of these transitions were observed simultaneously with the mm-RRL transitions and are available from the IRAM and Mopra data sets.

\begin{figure}
\begin{center}
\includegraphics[width=0.48\textwidth]{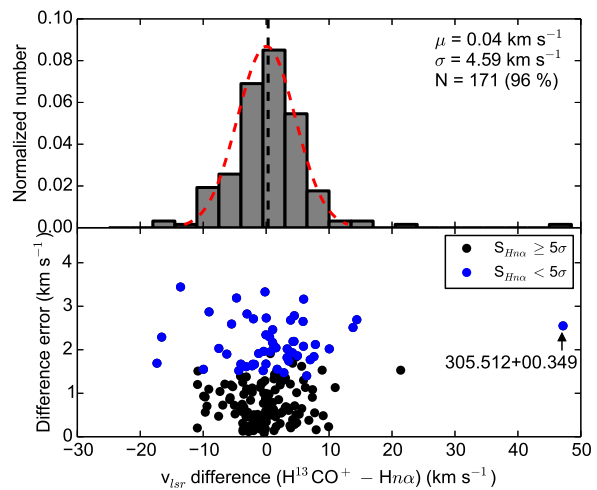}
\caption{\label{fig:vlsr_difference} Upper panel: Histogram of the v$_{lsr}$ difference between \hco\ and H$n\alpha$ RRL. The bin size is 3.5 \kms. Lower panel: Distribution of peak velocity errors with respect to the v$_{lsr}$ differences. The blue dots represent sources with weak mm-RRLs below 5\,$\sigma$. }
\end{center}
\end{figure}

As seen in all of the examples given in Figs.\,\ref{fig:stacked} and \ref{fig:2_spec}, the peak velocity of the mm-RRL emission is similar to that of the thermal molecular transition, which indicates that the two emission features are likely to be associated with the same clump. 
Figure\,\ref{fig:vlsr_difference} shows the differences between H$n\alpha$ RRL and \hco\ velocities. The velocity distribution is reasonably well fitted by a Gaussian profile (red dashed line) with a mean and standard error of 0.04\,$\pm$\,0.35\,\kms. This result is similar to that found in other cm- and mm-RRL studies made toward other samples of compact \hii\ regions (e.g., \uchii; \citealt{churchwell2009}) and GLIMPSE IRAC 8\,\mum\ bright \hii\ regions \citep{anderson2014}. These studies found mean velocity differences in a range of $-$1.3\,\kms\ $\leq \Delta$ v$_{lsr} \leq$ $+$1.5\,\kms. The strong correlation between the velocities of the molecular and ionized gas supports the physical association between the two.

Although the correlation between the mm-RRL and \hco\ velocities is generally very good, there are some sources where the velocity difference is larger than 15\,\kms, which is more than three times  the standard deviation ($\sigma$, 4.6\,\kms). In particular, the mm-RRL velocity we measure for AGAL305.512+00.349 (ID: N82) is extremely offset with respect to the systematic velocity as  marked in the bottom panel of  Fig.\,\ref{fig:vlsr_difference}. We note that in the majority 
of these cases the mm-RRL emission has a low S/N (see bottom panel of  Fig.\,\ref{fig:vlsr_difference},  blue dots) with the exception of AGAL301.136$-$00.226 (ID: N73), which is discussed in Sect.\,\ref{sec:hchii}.

\subsection{Relation between the molecular clouds and the \hii\ regions}
\label{sec:line-width}

\begin{figure}
\begin{center}
\includegraphics[width=0.48\textwidth]{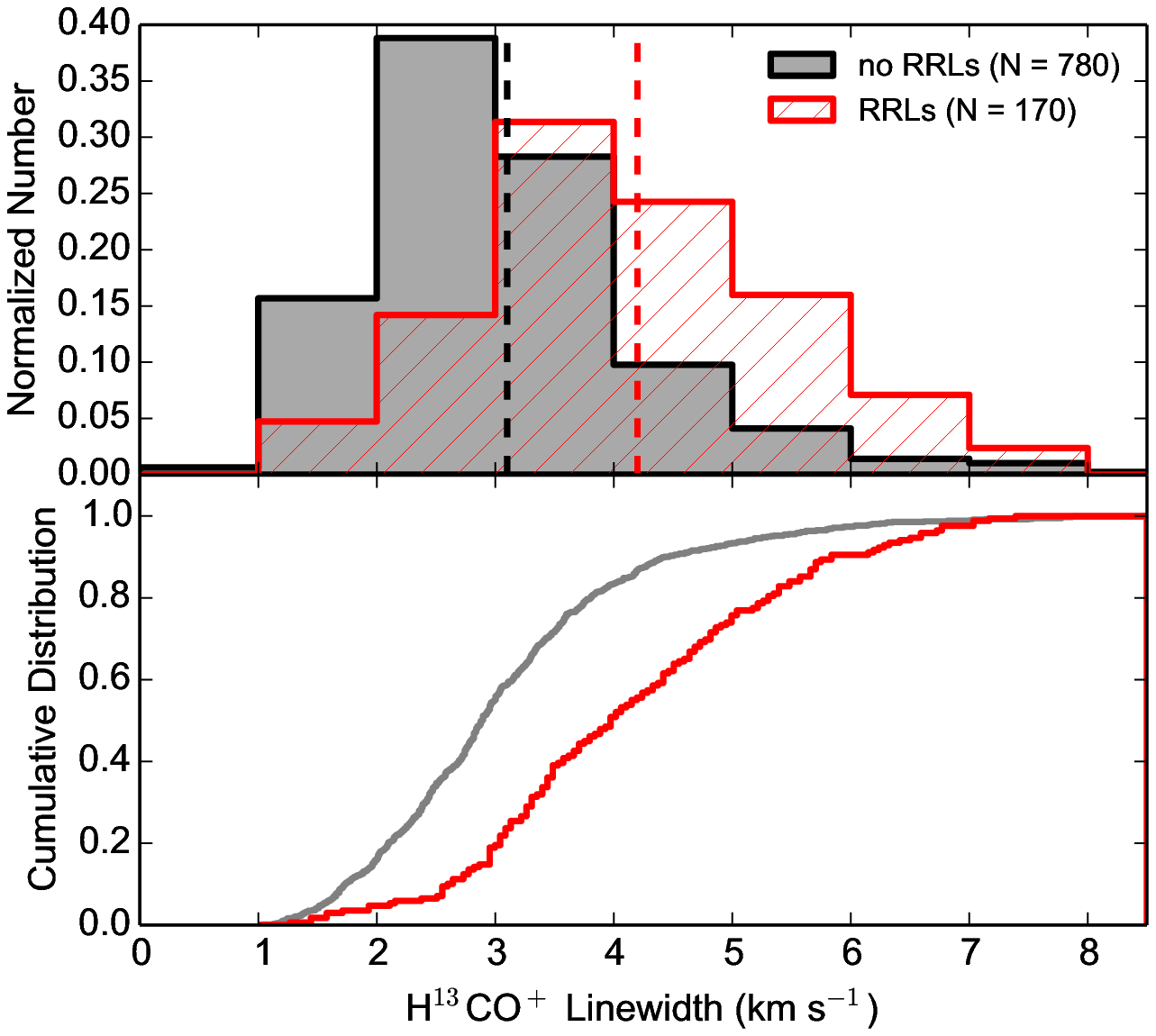}
\caption{\label{fig:h13co} Distributions of \hco\ linewidth for mm-RRLs associated (red) and mm-RRLs unassociated (gray) sources. Upper panel: Normalized histograms for the sources. The dashed lines represent mean values for each histogram. The bin size is 1.0\,\kms. Lower panel: Cumulative distribution functions.}
\end{center}
\end{figure}

To investigate whether there is a dynamic link between the \hii\ regions and the associated molecular clumps, we  compared the distribution of the mm-RRL and \hco\ linewidths, and no correlation was found (the Spearman's correlation coefficient, $\rho$, is 0.09). In Fig.\,\ref{fig:h13co}, however, histograms and cumulative distributions of clumps with and without mm-RRL detection show a difference between the \hco\ linewidths for the two different samples. The  linewidths of \hco\ toward clumps with mm-RRL detection (red curve) are notably broader than those toward clumps without a mm-RRL detection (gray curve), with mean linewidths of 4.18\,$\pm$\,0.11 (red dashed line) and 3.08\,$\pm$\,0.04\,\kms\ (black dashed  line) for clumps with and without mm-RRL detection, respectively. A \KS\ (KS) test shows that the null hypothesis that the two samples  are drawn from the same parent population is rejected with a $p-$value $\ll$ 0.001, implying that the existence of \hii\ regions traced by the mm-RRL detection contributes to the turbulence of the associated molecular clumps. 
This trend of molecular linewidth broadening with clump evolution is also confirmed by the results of \cite{wienen2012} and \cite{urquhart2013b}; both of these studies find a mean linewidth of the \nhthree\ transition (2.8\,\kms) toward ATLASGAL clumps associated with radio continuum sources, which is significantly broader than that found toward clumps without an embedded mid-IR sources (a mean linewidth of 1.9\,\kms).

\begin{figure*}
\begin{center}
\includegraphics[width=0.32\textwidth]{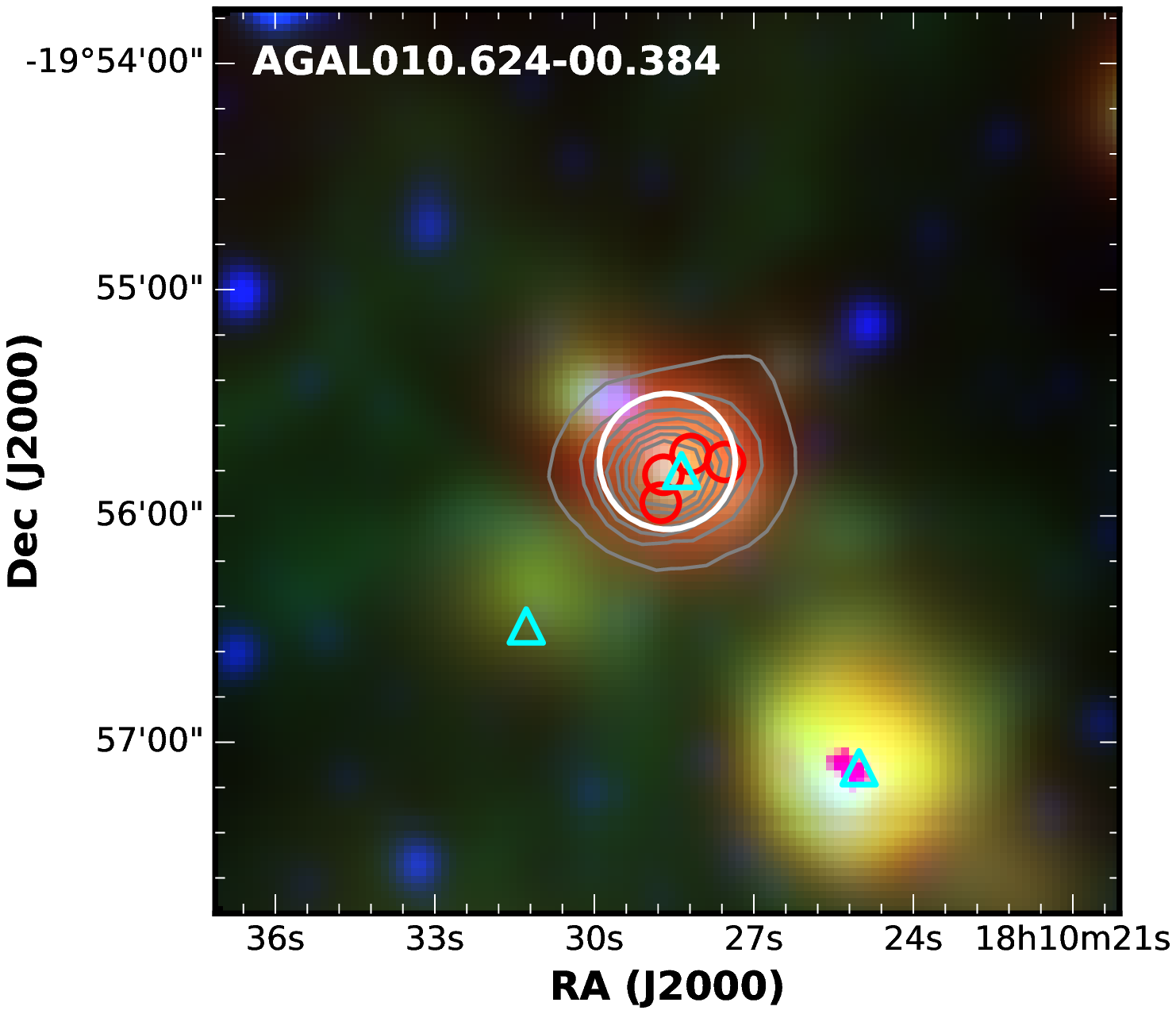}
\includegraphics[width=0.31\textwidth]{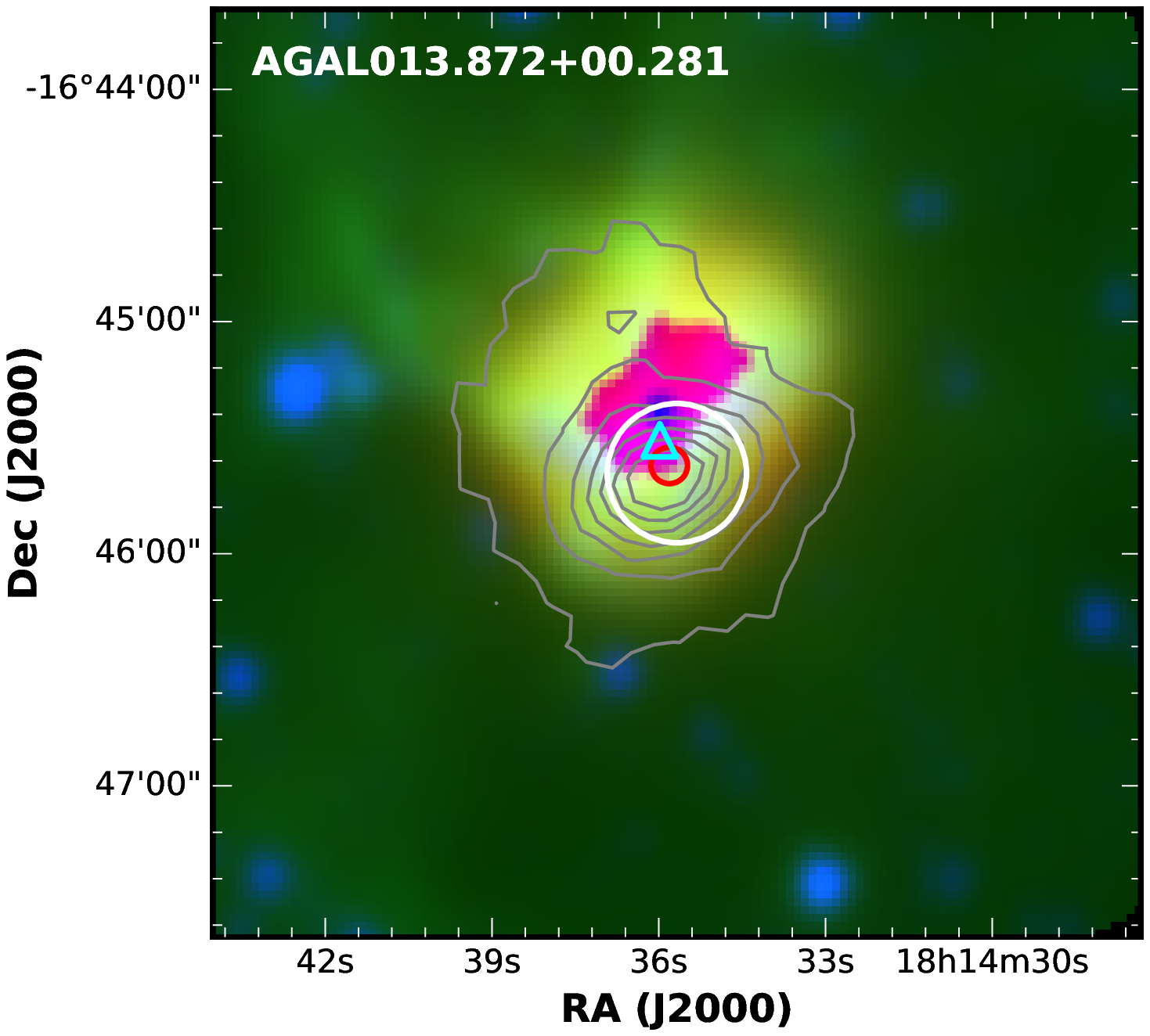}
\includegraphics[width=0.31\textwidth]{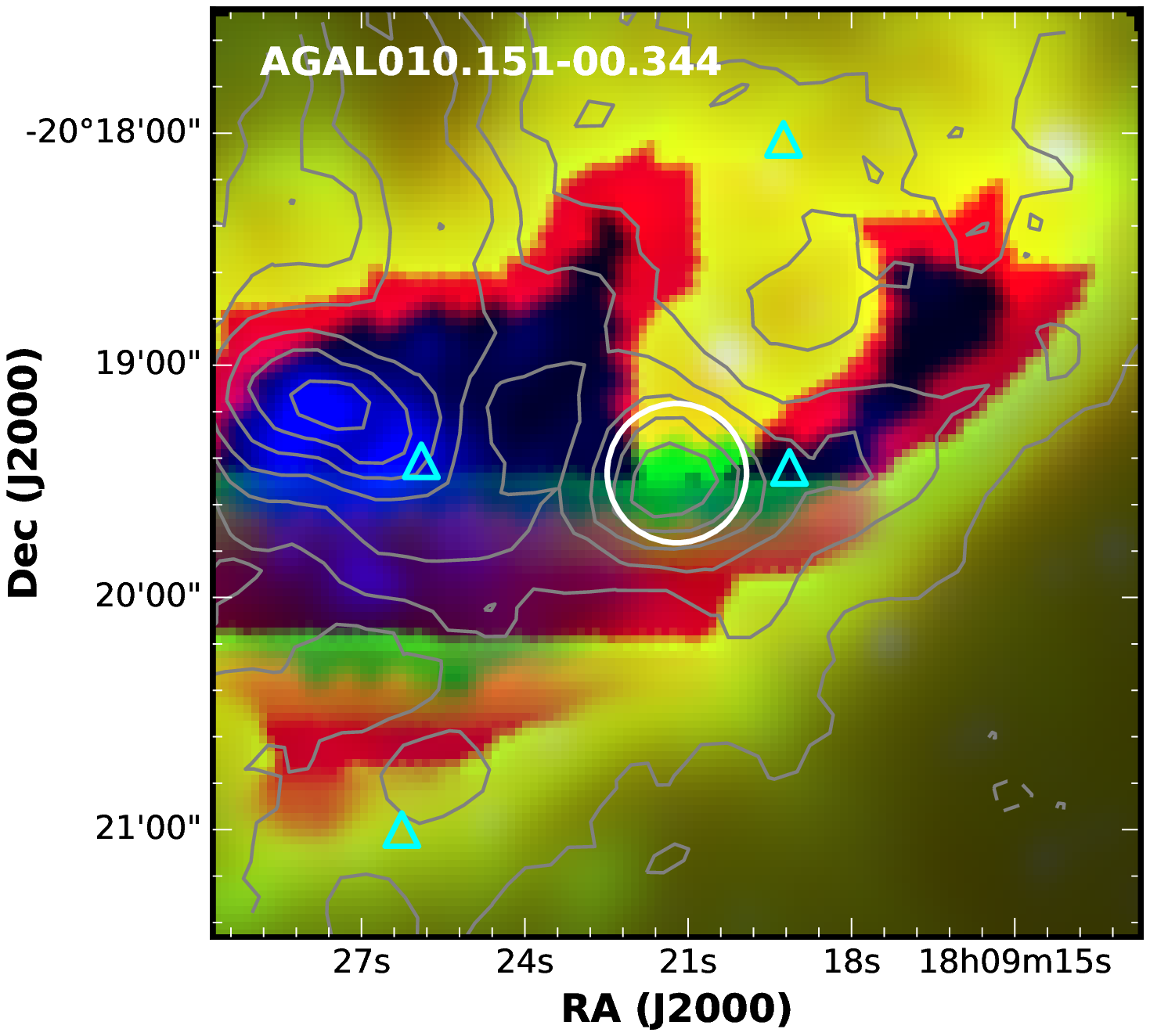}
\includegraphics[width=0.32\textwidth]{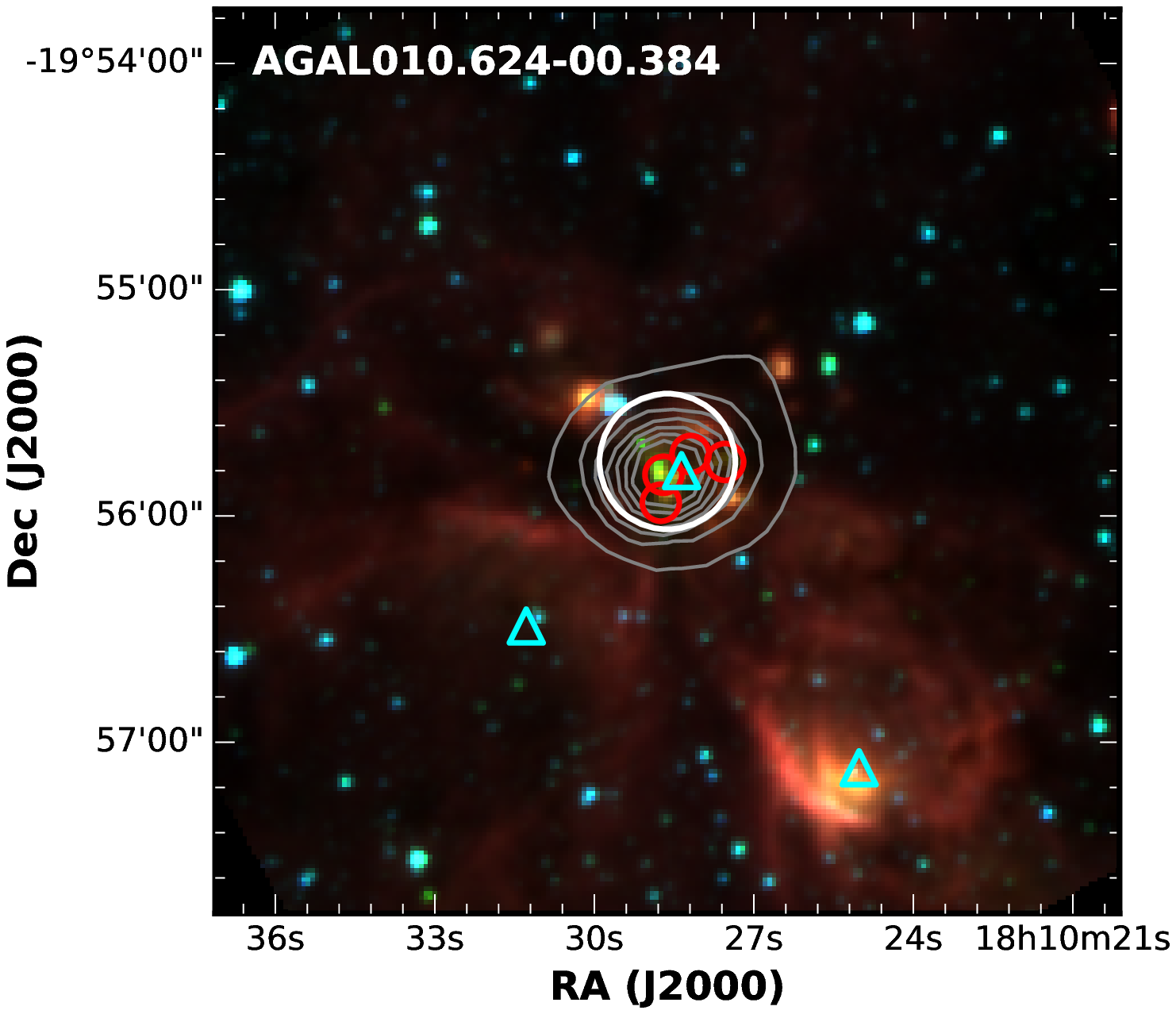}
\includegraphics[width=0.31\textwidth]{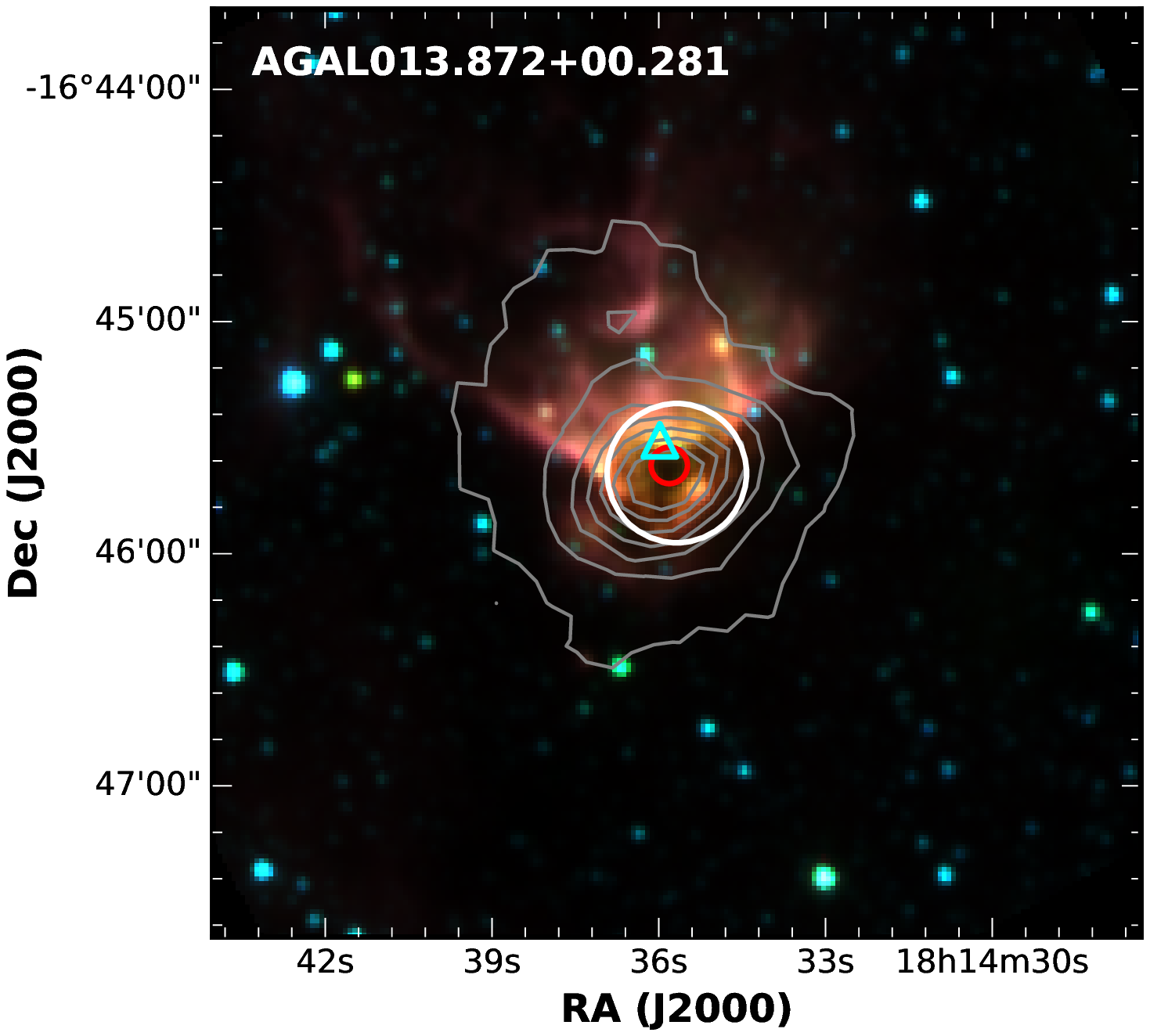}
\includegraphics[width=0.31\textwidth]{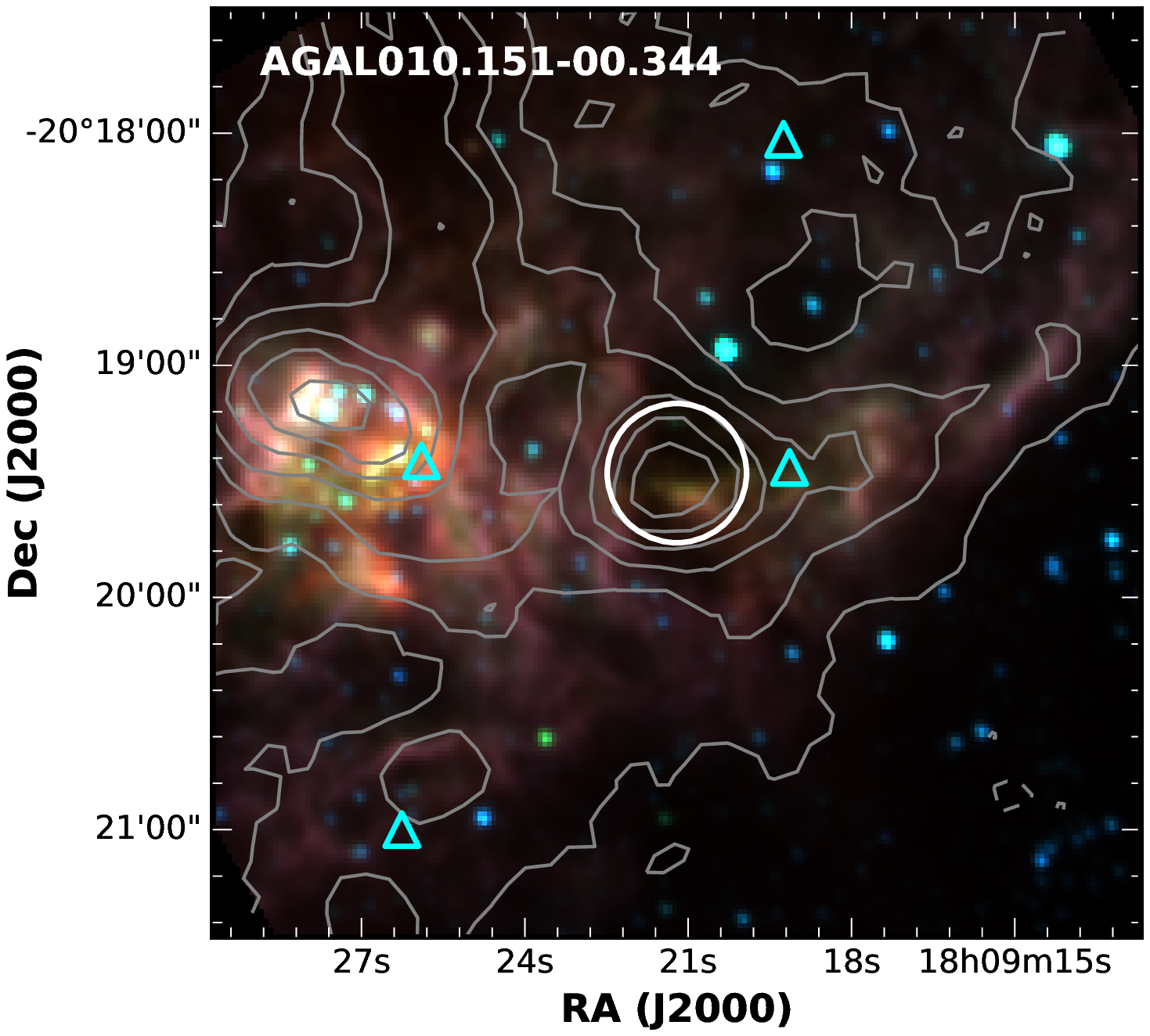}
\caption{\label{fig:midir} Clumps  representative of the three typical  mid-IR morphologies discussed in the text (from left to right: compact, extended, and complex.)  Upper panel: WISE three-color composite images (4.6 (blue), 12 (green), and 22\,\mum\ (red)). Lower panel: GLIMPSE IRAC three-color composite images (3.6 (blue), 4.6 (green), and 8\,\mum\ (red)). Gray contour lines indicate 870\,\mum\ dust continuum emission from the ATLASGAL survey. The red circles and cyan triangles are 5\,GHz CORNISH sources and 1.4\,GHz NVSS sources. 
The white circles are centered on the pointing positions used for our Mopra and IRAM observations and their diameters approximate the FWHM beams size of these telescopes in the 85--100 GHz range. The ATLASGAL beam has a size of $19''$ FWHM. }
\end{center}
\end{figure*}

\subsection{Radio continuum and mid-IR counterparts}
\label{sec:counterparts}

To further investigate the origin of the mm-RRLs, 
we  examined mid-IR images and searched for embedded radio continuum sources coincident with the clumps. For example,  because of its compactness a \hchii\ region only emits  weak radio emission that is optically thick throughout the radio range (i.e., $\nu$ $<$ 50\,GHz where its flux density varies as $\nu^2$). Given this weak emission, they might not have been detected in past shallow radio surveys that were conducted at longer cm wavelengths.

We  extracted mid-IR maps from the WISE and GLIMPSE surveys and created three-color images centered on the observed position. The mid-IR three-color images of three sources that are representative of compact, extended, and complex mid-IR emission
are presented in Fig.\,\ref{fig:midir}. These images provides a useful way to identify embedded protostellar objects and to investigate their local environments. We searched WISE point source catalogs \citep{cutri2012} and identified mid-IR counterparts within a radius of 18\arcsec\ (approximately the size of the ATLASGAL telescope beam) for 118 clumps\footnote{WISE point sources with the following quality flags  "D", "P", "H", or "O" are likely to suffer from contamination or confusion and are not reliable,  and so have been excluded.}. However, visual inspection of the three-color images reveals that many of the maps are affected by saturation indicating the presence of very bright sources within them. In total, we find mid-IR emission toward 127 clumps (including some sources that are saturated and that are not included in the WISE catalogs). We also  find that the saturation in the maps toward 43 sources is so bad that they cannot be used to evaluate the distribution of the mid-IR emission; many of these are found toward well known star-forming regions such as M17 and G305  (\citealt{povich2007} and \citealt{hindson2012}, respectively). We find no significant mid-IR emission associated with eight clumps.

We  also utilized several large continuum surveys at wavelengths of 3, 6, and 20\,cm  \citep{zoonematkermani1990,becker1994,condon1998,walsh1998,giveon2005,white2005,helfand2006,urquhart2007,urquhart2009,purcell2013} to identify embedded compact and \uchii\ regions.  
Using the same radius (18$''$) used to search for WISE counterparts, we  identified 116 compact radio sources that are positionally coincident with clumps associated with mm-RRLs. 

There is a possibility that radio sources other than \hii\ regions are the origin of the radio continuum, in particular extragalactic sources. The radio sources considered here are associated with the peak of the dust emission and are often extended, which is very unlikely for extragalactic sources. In addition, the extragalactic source counts are much lower at 5 GHz than at 1-2 GHz and hence the number of extragalactic sources, and the possibility of chance alignments with dust, is very low \citep{urquhart2013b}. \hii\ regions, even smaller \uchiis, clearly show counterparts in mid-IR data (e.g., GLIMPSE data), whereas extragalactic sources are not normally associated with any mid-IR emission \citep{hoare2012,purcell2013,urquhart2013b}. Since  all of the matched radio sources have an association with mid-IR emission, they can be classified as \hii\ regions.

In addition to the above-mentioned radio continuum surveys, 13 more clumps with mm-RRL detection have a 18\,GHz radio counterpart in the AT20G survey \citep{ricci2004}. In spite of the poor resolution and positional accuracy of the AT20G survey, these sources are associated with the clumps and mm-RRLs and so are likely to be \hii\ regions. Radio counterparts  for  five other clumps are also found in the literature \citep{kuchar1997,hindson2012,hindson2013}.
In total, compact radio emission is found to be associated with $\sim$75\% of the whole mm-RRL associated sample.
The positions of these radio sources are indicated in the images presented in Fig.\,\ref{fig:midir}. 
All of the radio sources are associated with clumps that are also associated with mid-IR emission, although ten of these are badly saturated in WISE 12 and 22 $\mu$m maps, but are recognizable in the GLIMPSE IRAC maps without any saturation. 
The positional correlation of the mid-IR source, the radio continuum emission, and the detection of the mm-RRL emission in 134 cases is consistent with the hypothesis that the ATLASGAL clumps are harboring compact \hii\ regions.

\begin{figure}
\begin{center}
\includegraphics[width=0.5 \textwidth]{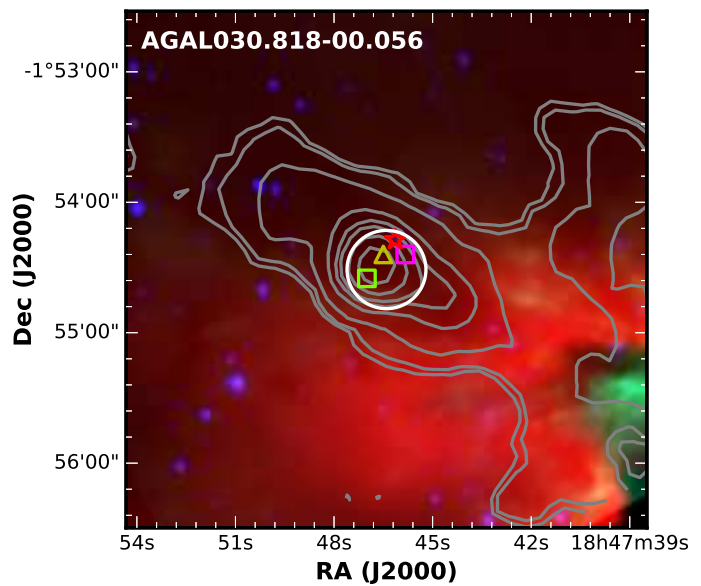}
\caption{\label{fig:hchii} Potential \hchii\ region candidate  W43-MM1. It is one of the bright dust continuum peak of the W43 ``mini-starburst'' region \citep{sridharan2014}.  Mid-IR three-color composite images (GLIMPSE IRAC 3.6\,\mum\ (blue), 8\,\mum\ (green), and WISE 22\,\mum\ (red)). The gray contours trace the 870\,\mum\ dust continuum emission. The white circles are centered on the pointing position used for IRAM observation. The red star symbol indicates a WISE 22\,\mum\ point source. The yellow triangle is an OH maser \citep{szymczak2004}. The green and purple squares indicate Class I and II \methanol\ masers \citep{larionov2007,szymczak2012}.  
}
\end{center}
\end{figure}

We are left with 44 clumps toward which mm-RRL emission is detected, but which are not associated with a compact radio source. There are three possible explanations for these sources: 1) the mm-RRL emission is coming from a nearby evolved \hii\ region, 2) the mm-RRL is associated with an optically thick \hii\ region at cm wavelengths whose flux density is below the limits of the exiting  cm-wavelength surveys, or 3)  the clump has not been included in any high-resolution radio continuum surveys.

Almost certainly, the 127 clumps with a compact mid-IR source are undergoing star formation. Conversely, the absence of a potential embedded source suggests that the mm-RRL emission is likely to be associated with nearby \hii\ regions. In addition, we have already found  large differences between the velocity of mm-RRL and the systemic velocity of the molecular gas for a number of associations which might be due to such contamination (Sect.\,\ref{sect:velocity_comparison}).
As previously mentioned, no mid-IR emission is detected toward eight of these clumps, one of which (N82) has a significant difference between the velocity of the mm-RRL and the systemic velocity of the clump, as have been marked in Fig.\,\ref{fig:vlsr_difference}, and it is therefore  likely that this association is due to contamination. We note that 3 of the 134 compact radio sources (N59,
N73, N104) are matched to mm-RRL emission where the velocities of the mm-RRLs with respect to the molecular gas are larger than 15\,\kms.
Therefore, they are likely unassociated with molecular clumps and the result of contamination from nearby \hii\ regions. 
Nevertheless, they also have compact radio and mid-IR sources.  It means that we cannot fully exclude the possibility that the large velocity difference is caused by other reasons such as internal turbulent motions of the \hii\ regions. 

Owing to the relatively large size of the Mopra and IRAM beams, it is possible that some of the mm-RRL detections originate from large-scale \hii\ regions in close proximity to a significant number of clumps and not the star-forming regions associated with the ATLASGAL clumps. Bright and extended \hii\ regions might even be picked up by the sidelobes of the telescopes.

The radio counterparts found for the northern hemisphere clumps are drawn primarily from the CORNISH survey which is an unbiased 5\,GHz radio continuum survey (\citealt{hoare2012}), and so the third possibility that a particular clump has not been included in any high-resolution radio continuum surveys can be discarded. In this region of the plane we find 11 mm-RRL detections not associated with a compact radio source. Of these, nine are located near the intense star-forming regions M17, G34, W49, and W51. Their mid-IR images are almost completely saturated in the WISE 22\,\mum\ band and show complex GLIMPSE IRAC 8\,\mum\ emission. The non-detection of any compact radio continuum and the close proximity to star-forming complexes lead us to conclude that the mm-RRL emission is due to contamination and is not associated with the clumps. 
The remaining two sources, AGAL029.911$-$00.042 (ID: N30) and AGAL030.818$-$00.056 (ID: N37; see also Fig.\,\ref{fig:hchii}), are associated with compact mid-IR emission. Source
N30 is related to SiO molecular emission with high-velocity line wings, which indicates the presence of outflows from deeply embedded MYSOs \citep{csengeri2016_sio}. Source N37 is associated with masers (e.g., Class I \& II methanol and OH masers) that also indicate an earlier stage than an evolved \hii\ region. Therefore, these two sources are good potential \hchii\ region candidates.

The radio associations identified for clumps located in the southern hemisphere have largely been drawn from targeted observations of an MSX color selected sample of MYSOs (\citealt{urquhart2007}) and methanol masers (\citealt{walsh1997}). A consequence of this is that not all mm-RRL emitting clumps have been observed. Therefore, the detection of a mm-RRL may be revealing the location of \hii\ regions that have not previously been discovered at radio wavelengths.  There are 25 mm-RRL sources detected in the fourth Quadrant that have not previously discussed; 19 are located near the edges of large complexes where mm-RRL emission is likely to be the result of contamination from evolved \hii\ regions; however, the remaining 6 detections are all associated with compact mid-IR emission and are likely to be good \hii\ region candidates. 

\begin{figure}
\begin{center}
\includegraphics[width=0.46\textwidth]{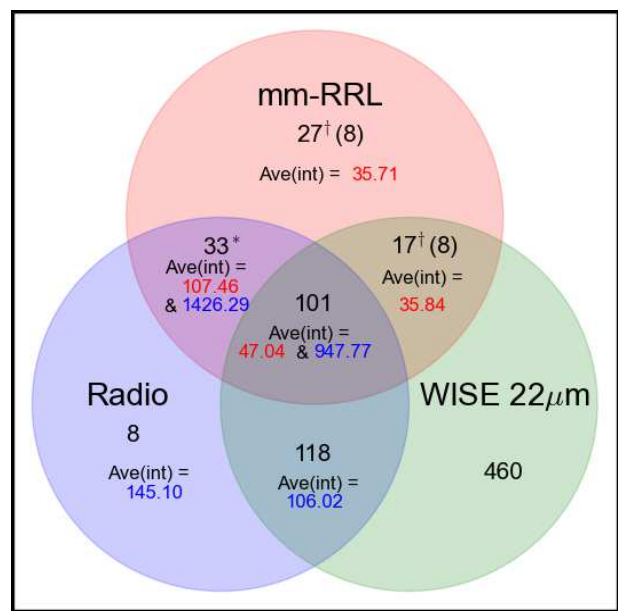}
\caption{\label{fig:venn_diag} Venn diagram illustrating the distributions of ATLASGAL clumps (black numbers) that are in association with mm-RRL detection, radio continuum sources (only considering large continuum surveys for clumps without mm-RRL detection, while the number of radio sources with mm-RRL is based on complete, individual searches), and WISE 22\,\mum\ emission (only considering association with WISE 22\,\mum\ point sources). In addition, average integrated fluxes of total H$n\alpha$ and 6\,cm radio continuum are given as red and blue numbers in Jy\,\kms\ and mJy, respectively. The star  ($^{*}$) indicates that these 33 sources are saturated in WISE 22\,\mum\ emission and the dagger  ($^{\dag}$) indicates that contaminated sources have been included. The numbers in parentheses  represent 8 mid-IR dark sources among the 27 clumps with only mm-RRL and 8 potential \hii\ candidates among the 17 clumps with both the mm-RRL and the WISE 22\,\mum\ point source. }
\end{center}
\end{figure}
Figure\,\ref{fig:venn_diag} visualizes in a Venn diagram the association of the mm-RRL detected clumps with radio and infrared WISE 22\,\mum\ emission.
It also provides average values of integrated 6\,cm continuum (blue text) and mm-RRL (red text) intensities.
In summary, we  identify 44 sources that show mm-RRLs and are not matched to a compact radio source. Of these, 28 are located near some of the most intense star-forming regions in the Galaxy where the nature of the emission cannot be reliably determined.
In addition, eight clumps are found to be mid-IR dark and therefore unlikely to harbor \hii\ regions; in all of these cases there is evidence of a nearby \hii\ region that is likely to be the source of the observed mm-RRL emission.
Nevertheless, the fact that a mid-IR dark clump may contain an \hii\ region cannot be completely excluded in order to explain the mm-RRL detection toward the mid-IR dark clump. 
Finally, there are eight clumps that are associated with bright mid-IR emission,  many of which are also associated with methanol masers; these are considered to be good \hii\ region candidates.

\section{Analysis and discussion }\label{sec:analysis}
\subsection{Comparison of the mid-IR and radio properties}
\label{sec:comparison_midir_radio}

\begin{figure}
\centering
\includegraphics[width=0.5\textwidth, trim = 0 10 0 20]{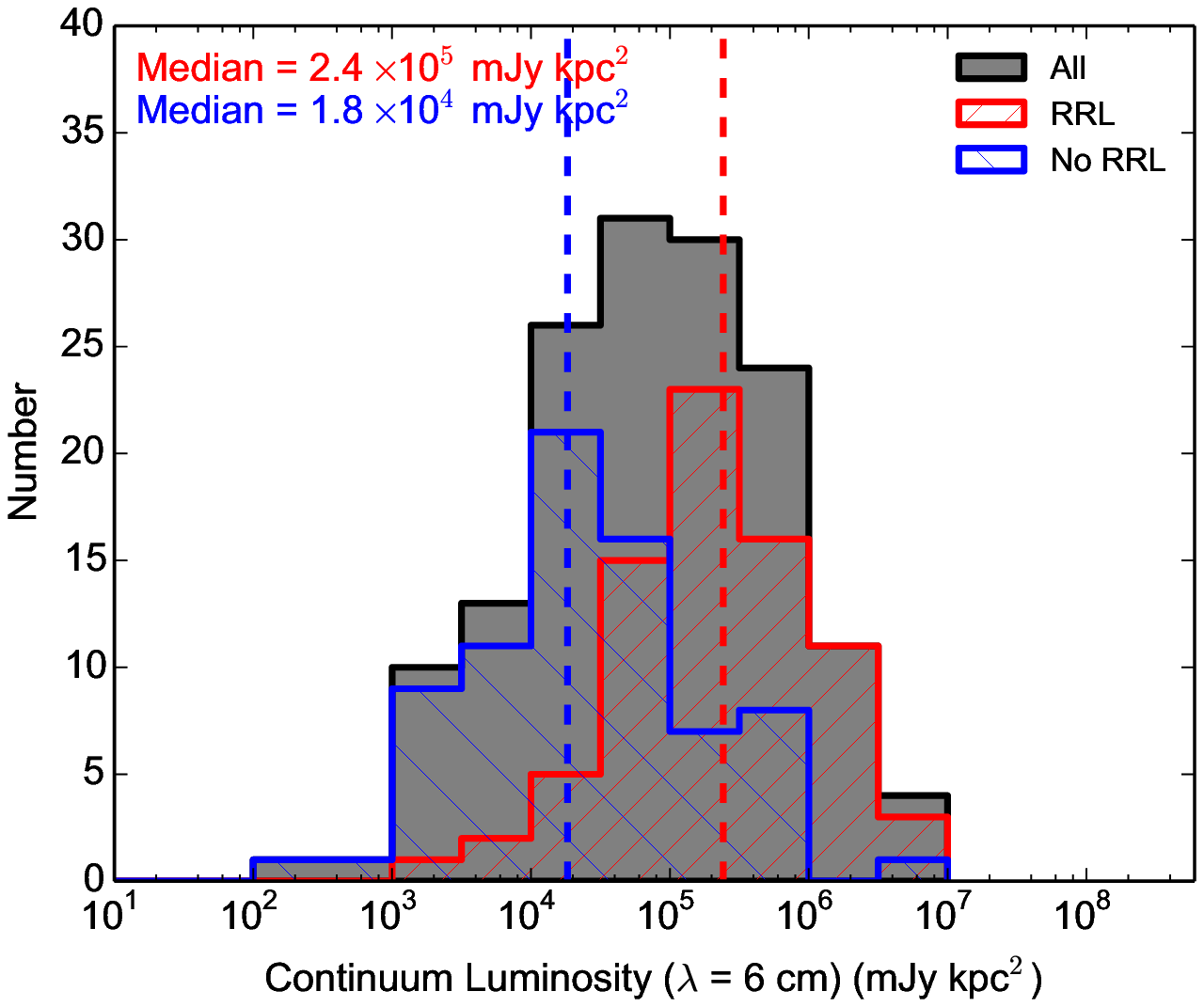}
\includegraphics[width=0.5\textwidth, trim = 0 20 0 10]{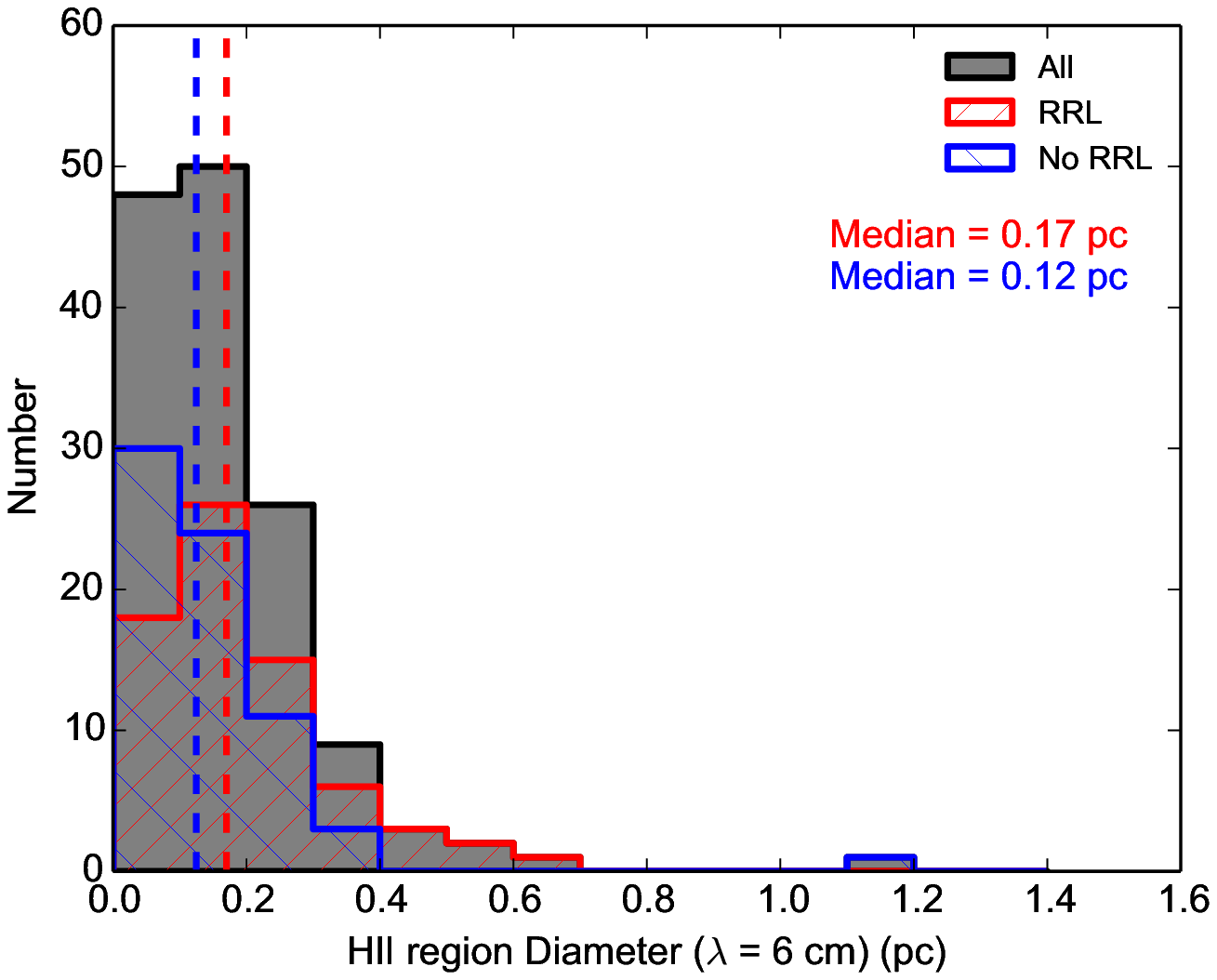}
\caption{\label{fig:radio_rrl_nonrrl} Upper panel: Radio continuum luminosity for sources with and without mm-RRL emission. The bin size is 0.5\,dex. Lower panel:  \hii\ region diameters. The bin size is 0.1\,pc. The blue and red dashed lines indicate median values of the radio continuum luminosity and \hii\ region diameter for mm-RRL and non mm-RRL sources. The gray histograms of both plots show the distributions of the full sample}.
\end{figure}

In the previous section we discussed the radio and mid-IR counterparts that are associated with clumps with mm-RRL detection. We found that the vast majority of the mm-RRL emission sources are associated with strong mid-IR emission and compact radio emission sources and these are therefore associated with compact \hii\ regions. The full sample of 976 observed ATLASGAL clumps also includes clumps associated with radio and mid-IR emission but toward which no mm-RRL emission has been detected. 
In this section we  compare the mid-IR and radio properties of the \hii\ region with and without detectable mm-RRL emission in an effort to understand the differences between the two samples. 
\begin{figure}
\centering
\includegraphics[width=0.5\textwidth, trim = 0 10 0 20]{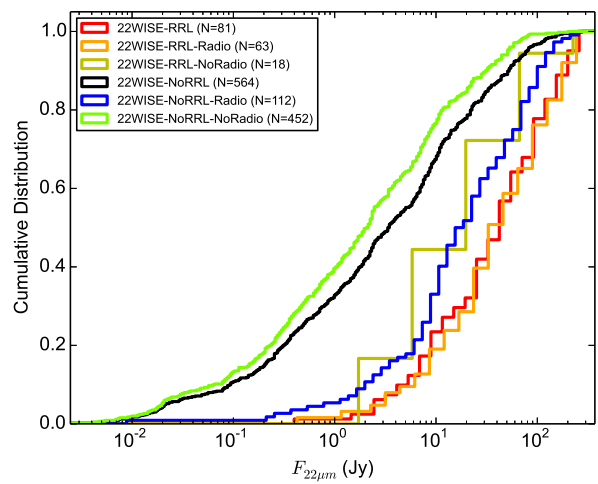}
\caption{\label{fig:wise_rrl_radio} Cumulative flux distributions of WISE 22\,\mum\ point sources. The colors indicate associations with a detection of a mm-RRL and a radio continuum counterpart. WISE 22\,\mum\ point sources with fluxes over 330\,Jy are excluded as the detectors saturate and the fluxes are unreliable.}
\end{figure}

Compact radio sources have been matched to 242 clumps (only considering the main large radio continuum surveys that we used when searching for radio counterparts) from the whole sample of 976 clumps ($\sim$25\% of the sample), but mm-RRL emission is only detected toward slightly less than half of these. 
In Fig.\,\ref{fig:radio_rrl_nonrrl} we compare the 6\,cm continuum luminosity and \hii\ region diameter of radio sources with and without mm-RRL emission association.
The information regarding radio sources, parental clumps, and heliocentric distances is provided in Table\,\ref{tb:wise_radio}.

It is clear from the upper panel of Fig.\,\ref{fig:radio_rrl_nonrrl} that the radio sources associated with mm-RRL emission are significantly brighter than those without. 
Such a clear difference is not found for the \hii\ region diameters, which are similar for both source groups. This suggests that the mm-RRL detectability is mainly sensitive to the radio continuum brightness of the \hii\ regions.
We note that we only use the 6\,cm continuum detections for the analysis presented here and the rest of the paper, as this data is available for the majority of the clumps and all of the surveys have a similar angular resolution and sensitivity leading to a consistent data set to study the continuum association. 

\begin{table*}
\begin{center}
\caption{\label{tb:wise_radio} WISE 22\,\mum\ point sources and radio continuum sources matched with mm-RRLs detection.}
\begin{tabular}{c c . . . . .}
\hline
 && \multicolumn{3}{c}{6\,cm radio continuum} & \multicolumn{1}{c}{WISE 22\,\mum} & \\
 \hline
No. && \multicolumn{1}{c}{S$_{\rm 6\,cm}$} & \multicolumn{1}{c}{$\int$S$_{\rm 6\,cm}$} & \multicolumn{1}{c}{Angular diameter}  & \multicolumn{1}{c}{$F_{\rm 22\,\mu m}$$^{\dag}$}  & \multicolumn{1}{c}{Distance}\\
 && \multicolumn{1}{c}{(mJy/beam)} & \multicolumn{1}{c}{(mJy)} & \multicolumn{1}{c}{(\arcsec)} & \multicolumn{1}{c}{(Jy)} & \multicolumn{1}{c}{(kpc)} \\
\hline\hline
1 && 667.5 & 795.0 & 3.6 & 12.3 & $--$ \\
4 && 13.8 & 170.3 & 19.0 & $--$ & 3.5 \\
5 && 34.6 & 57.7 & 2.2 & $--$ & 8.6 \\
6 && 360.6 & 2060.9 & 4.9 & $--$ & 4.9 \\
7 && 109.1 & 196.0 & 1.6 & 17.2 & 13.7 \\
8 && 21.7 & 50.2 & 5.2 & 28.6 & 14.4 \\
9 && 163.9 & 1155.9 & 5.7 & 186.9 & 4.0 \\
10 && 72.2 & 207.9 & 2.4 & 4.2 & 13.6 \\
11 && 287.9 & 12616.4 & 16.2 & $--$ & 2.4 \\
12 && 40.7 & 946.8 & 8.2 & 38.8 & 4.6 \\
13 && 10.5 & 603.9 & 19.1 & 59.3 & 1.9 \\
14 && 24.7 & 1447.6 & 15.4 & $--$ & 4.4 \\
21 && 97.7 & 519.3 & 3.1 & 40.9 & 17.1 \\
23 && 48.0 & 1277.9 & 14.6 & $--$ & 3.0 \\
24 && 121.6 & 342.1 & 2.3 & 31.1 & 12.1 \\
25 && 86.4 & 510.2 & 8.6 & $--$ & 4.5 \\
27 && 33.0 & 49.0 & 1.5 & $--$ & 9.1 \\
\hline
\end{tabular}
\tablebib{{6\,cm radio continuum emission:~\citet{becker1994,white2005,urquhart2007,urquhart2009,purcell2013}. WISE 22\,\mum\ point source}: ~\citet{cutri2012}.  Distance:~\citet{urquhart2014_atlas, csengeri2016_sio}.}
\tablefoot{{A portion of the entire table is given here for simplicity. The entire table is available at the CDS. {$^{(\dag)}$}The $F_{22\,\mu m}$ is a measured flux of a point source from the WISE point source catalog, which is the closest one without a bad flag within a searching radius of 18$''$.} WISE 22\,\mum\ point sources with fluxes over 330\,Jy are excluded as the detectors saturate and the fluxes are unreliable.
Regarding the displayed values of 6\,cm radio continuum emission, explanations of how to determine the values are available in  Sects.\,\ref{sec:radio_rrl_intensity} and \ref{sec:em_ne}.}
\end{center}
\end{table*}


Within a search radius of 18$''$ at a observed position, WISE 22\,\mum\ point sources are matched to 696 clumps in the sample ($\sim$70\%) with 118 being associated with mm-RRL detections and 578 associated with mm-RRL quiet clumps.  
This would suggest that the majority of the sources observed are actively forming stars, with approximately 35\% (242/696) of these already harboring young stellar objects and/or compact \hii\ regions. Fig.\,\ref{fig:wise_rrl_radio} shows the cumulative flux distributions\footnote{The 22\,\mum\ magnitudes were converted to flux in Jy units. A zero point magnitude of 8.2839, a color correction factor of 1.0 for 22\,\mum,\ and an additional correction factor of 0.9 due to an uncertainty in the calibration were used to calculate the conversion factor \citep{Wright2010}.} of the WISE 22\,\mum\ counterparts for various associations with the mm-RRL detection and radio continuum counterpart. 
The WISE point sources without the bad flags are valid, but their fluxes brighter than 330\,Jy are not reliable (see \citealt{csengeri2014} for more details). 
Therefore, the plots for fluxes of the WISE 22\,\mum\ counterparts have only used the WISE 22\,\mum\ point sources with a flux less than 330\,Jy. 
It is clear that the mm-RRL samples (red and orange curves) are associated with the brightest WISE sources and the flux distributions are significantly different from WISE sources not associated with a mm-RRL (black curve). The distribution of the radio and mid-IR associated clumps (blue curve) is significantly different from both the radio quiet clumps and the radio loud and mm-RRL associated clumps (the KS test is able to reject the null hypothesis that these are drawn from the same parent population with a $p-$value $\ll$ 0.001).

\begin{figure}
\includegraphics[width=0.5\textwidth, trim = 0 10 0 20]{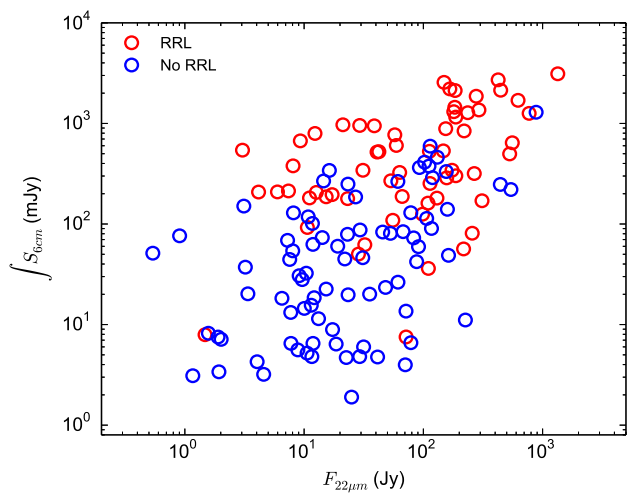}
\caption{\label{fig:flux_ratio_raio_wise} Integrated 6\,cm radio continuum flux as a function of 22\,\mum\ flux for the mm-RRL quiet and bright radio sources.}
\end{figure}

We have previously shown that radio sources not associated with mm-RRL emission tend to have weaker continuum luminosities. The distribution of their mid-infrared fluxes shown in Fig.\,\ref{fig:wise_rrl_radio} reveals that they also tend to have significantly lower mid-IR emission. Both the radio and mid-IR emission are distance dependent quantities and it is therefore possible that these trends are due to a physical difference in the properties of the embedded objects or due to a distance bias.
If these trends arise from physical differences in the embedded objects, we would expect to see separation between both radio source groups with mm-RRL and without, in radio and mid-IR flux distributions. To minimize the potential for bias we only use clumps associated with a single radio source in a searching radius; this ensures that the flux of the radio source is compared with a flux of the closest WISE point source and that the radio and mid-IR emission are related to the same ionizing star.
The WISE 22\,\mum\ fluxes of 25 point sources matched with the radio sources are severely saturated. Therefore, we used the MSX 21\,\mum\ flux for those saturated sources instead of excluding them from the radio and mid-IR flux distributions. Fig.\,\ref{fig:flux_ratio_raio_wise} shows a scatter plot of integrated 6\,cm radio continuum flux as a function of the WISE 22\,\mum\ point source flux in association with mm-RRL detection.
It is clear from this plot that the mm-RRLs are associated with the brightest radio sources and that this trend is largely independent of the mid-infrared emission. This is confirmed by a KS test ($p-$value $\ll$ 0.001). In other words, these two populations are likely to have different physical conditions and are not the result of a distance bias. In spite of the small $p-$value of the KS test, however, these two populations are not obviously separated in the distributions.

\begin{figure}
\centering
\includegraphics[width=0.5\textwidth, trim = 0 10 0 20]{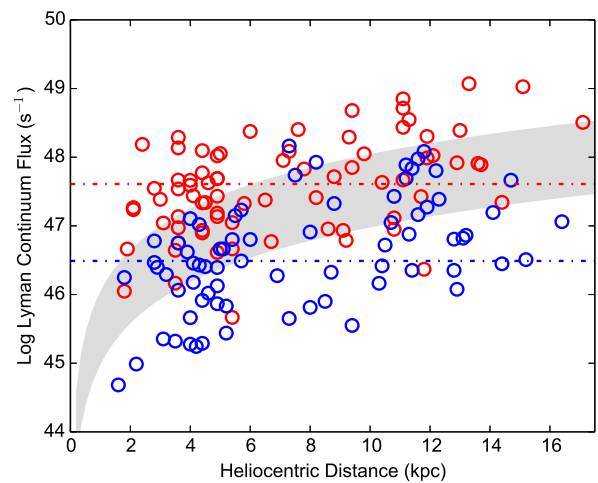}
\caption{\label{fig:lyman_dist} Lyman continuum flux as a function of heliocentric distance. The red and blue circles represent 6\,cm radio continuum sources associated with and without mm-RRL sources. The gray area is filled between Lyman continuum fluxes estimated with median 6\,cm integrated fluxes of radio sources associated with and without mm-RRL. The red and blue dot-dashed lines show median values in the Lyman continuum flux distributions for those sources, respectively.}
\end{figure}

Fig.\,\ref{fig:lyman_dist} shows the Lyman continuum flux as a function of the heliocentric distance for radio continuum sources associated with ATLASGAL clumps. In this plot the mm-RRL associated and mm-RRL quiet radio sources are indicated as red and blue circles, respectively. The Lyman continuum fluxes are calculated using

\begin{equation}
\left(\frac{N_{i}}{\rm{photon\,s^{-1}}}\right)=9\times10^{43}\,\,\left(\frac{\int S_{\nu}}{\rm{mJy}}\right)\,\, \left(\frac{D^{2}}{\rm{kpc}}\right) \,\, \left(\frac{\nu^{0.1}}{\rm{5\,GHz}}\right)
\label{eqn:ni}
,\end{equation}

\noindent where $\int S_{\nu}$ is the integrated radio flux density measured at frequency  $\nu$ and $D$ is the heliocentric distance to the source \citep{urquhart2013b}. This assumes the \hii\ regions are optically thin. This  significantly underestimates the Lyman continuum flux for more compact \hii\ regions, but this assumption is justified (see next two sections). Radio continuum sources with mm-RRL mostly show Lyman continuum fluxes of $\geq 10^{47}$\,photon\,s$^{-1}$, whereas the distribution of mm-RRL quiet radio sources tends to be associated with weaker \hii\ regions ($\sim 10^{46}$\,photon\,s$^{-1}$). The median of the Lyman continuum flux distribution for radio sources with mm-RRL (red dot-dashed line) is close to $10^{48}$\,photon\,s$^{-1}$, while the median of the Lyman continuum flux distribution for mm-RRL quiet radio sources (blue dot-dashed line) is less than $\sim 10^{46.5}$\,photon\,s$^{-1}$. 

The Lyman continuum fluxes of a large fraction of the radio sources with mm-RRL correspond well to the expected fluxes of late O-type stars. \hii\ regions ionized by O-type stars have Lyman continuum fluxes of $> 10^{48}$\,photon\,s$^{-1}$ and tend to have larger \hii\ region diameters than \hii\ regions associated with B-type stars that have fluxes $< 10^{48}$\,photon s$^{-1}$ (\citealt{urquhart2013b}). The sensitivity of the Mopra and IRAM observations appears to be limited to the detection of mm-RRL from the brightest late O-type stars.

\subsection{The recombination line and radio continuum emission}
\label{sec:radio_rrl_intensity}

\begin{figure}
\begin{center}
\includegraphics[width=0.49\textwidth, trim = 0 10 0 20]{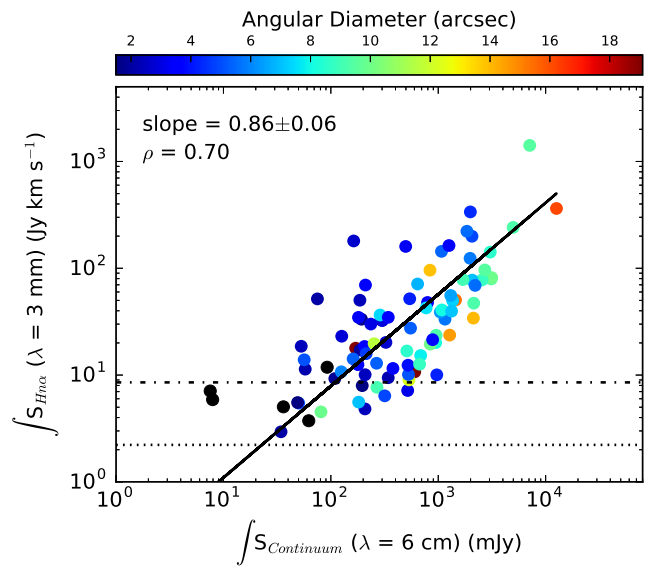}
\caption{\label{fig:rrl_cm} Integrated flux of radio continuum at 6\,cm versus integrated flux of mm-RRLs at 3\,mm. The black solid line indicates the best-fit line determined by a xy-bisector fit to the data. The black dot-dashed and dotted lines indicate 3\,$\sigma$ integrated fluxes for IRAM 30m (2.22\,Jy\,\kms) and Mopra 22m (8.54\,Jy\,\kms). The color bar represents an angular diameter of continuum source in units of arc second. The black filled-circles show no angular size for unresolved radio sources. The errors on the integrated fluxes of mm-RRLs are 1.82 Jy\,\kms\ on average.}  
\end{center}
\end{figure}

\begin{figure}
\centering
\includegraphics[width=0.49\textwidth, trim = 0 10 0 20]{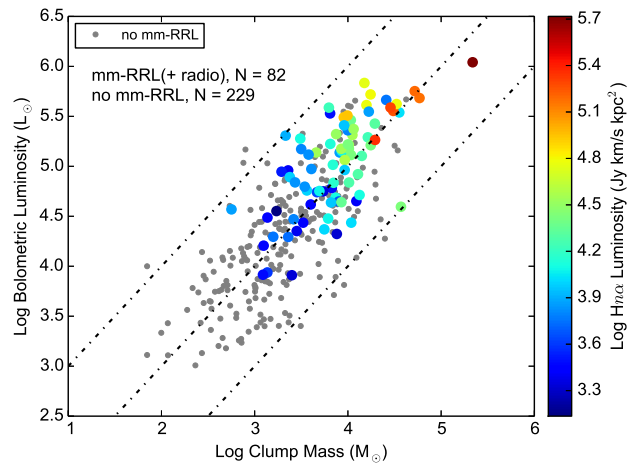}
\caption{\label{fig:lbol_clumpMass_lrrl} Bolometric luminosity as a function of clump mass for \hii\ regions presented in \cite{urquhart2014_atlas}. The colors represent the mm-RRL luminosity. The gray dots indicate clumps without mm-RRL detection.  
The lower, middle and upper diagonal dot-dashed lines represent the $L_{\rm bol}/M_{\rm clump}$ = 1, 10, and 100\,L$_{\odot}/$M$_{\odot}$, respectively. } 
\end{figure}

The radio continuum sources are smaller than the beam size of the Mopra and IRAM observations and so in some cases there are two or more radio sources located within the beam. It is thus likely that the observed mm-RRL emission has contributions from several radio continuum sources. To account for this we have summed the flux of all radio sources within the beam to obtain the total integrated radio flux. 
In Fig.\,\ref{fig:rrl_cm} we plot spatially integrated 6\,cm continuum flux density versus velocity-integrated mm-RRL flux. The mm-RRL detection thresholds are estimated from $\Delta w = \sigma \delta v \sqrt{N}$, where $\delta v$ is the channel width in units of \kms\ and $N$ is the number of channels over which the emission is found (we have used a value of 10). 
Comparing the continuum and line fluxes we find them to be strongly correlated (the Spearman's correlation coefficient, $\rho$, is 0.70 with a $p-$value $\ll$ 0.001). An xy-bisector fit to these data results in a slope of 0.86\,$\pm$\,0.06.

Given that the mm-RRLs and radio continuum emission are tracing the same volume of gas, we would expect to find a linear relationship if the \hii\ regions continuum emission were optically thin. However, if it were optically thick at 6\,cm we would expect the mm-RRL and continuum flux to deviate significantly from a linear relationship. The most compact \hii\ regions are the most likely to have optically thick emission (blue filled circles in Fig.\,\ref{fig:rrl_cm}); however, these regions do not deviate from the linear relationship. The fit to the fluxes shown in Fig.\,\ref{fig:rrl_cm} is close to linear, which suggests that all of the \hii\ regions are broadly optically thin.

In Fig.\,\ref{fig:lbol_clumpMass_lrrl} we show the relationship between the bolometric luminosity of all embedded \hii\ regions identified by the RMS survey (\citealt{lumsden2013}) in each clump \citep{urquhart2014_atlas} and the mass of the host clump. In this plot we also indicate the luminosity of the associated mm-RRL emission.
When we calculated the luminosities of the mm-RRLs for the clumps with mm-RRL detections, we only used a subsample of the clumps with 6\,cm radio continuum sources  in order to reduce potential contamination by nearby \hii\ regions unassociated with the clumps, which is more of an issue for the lower resolution 20 cm NVSS survey.

There is a strong correlation between these three parameters with a clear trend for more luminous \hii\ regions giving rise to stronger mm-RRL emission and being associated with more massive clumps. 
We evaluated a Spearman's correlation coefficient ($\rho$) for clumps with mm-RRL detection, without detection,  and all clumps and obtained 0.67, 0.70, and 0.76 with $p-$values $\ll$\,0.0001, respectively. 
These results are higher than the partial Spearman's correlation coefficient ($r=0.64$) of the MYSO, \hii\ region, and multiphase subsamples in \cite{urquhart2014_atlas}. 

The Spearman's correlation coefficient ($\rho=0.67$) of the clumps with mm-RRL detection is significantly higher than the partial Spearman's correlation coefficient ($r=0.53$) for the \hii\ region subsample which are CORNISH \hii\ regions of \cite{urquhart2013b}. The clumps with mm-RRL detection are not only associated with mm-RRLs, but also have compact radio continuum sources. 
The clumps are therefore a subsample of the CORNISH \hii\ regions with a brighter flux than those of \cite{urquhart2013b}. The fact that the mm-RRLs are associated with brighter radio continuum sources is also supported by the distribution seen in Fig.\,\ref{fig:flux_ratio_raio_wise}. 

Clumps without a mm-RRL detection (gray dots) show a broad distribution from  low to high masses and bolometric luminosities. 
On the contrary, clumps that are detected mm-RRLs (colored circles) tend to be associated with a large mass and bolometric luminosity. 
Similarly, the \hco\ linewidths increase in clumps with the mm-RRL detection, as was shown in Sect.\,\ref{sec:line-width}, although no clear correlation exists between the linewidth of \hco\ and mm-RRLs for the mm-RRL luminosity in Fig.\,\ref{fig:lbol_clumpMass_lrrl}. 
It seems that the clumps with mm-RRL detection are more massive and have larger \hco\ linewidths, the latter also implying that they are in virial equilibrium.

A possible reason for the lack of  correlation between the linewidths of mm-RRL and \hco\ may be that different size scales are involved; the linewidths of mm-RRL are linked to the small scales of the \hii\ regions, whereas the linewidths of \hco\ are averages over the larger scale molecular clumps. 
However, we found by comparing samples with and without \hii\ regions that the \hii\ regions detected by mm-RRLs show enhanced turbulent motions in their environments, either directly or indirectly.
This mm-RRL luminosity gradient and the \hco\ linewidths reveal that the most massive clumps that  host  \hii\ regions and that are associated with strong mm-RRLs also have  broader molecular linewidths.

\begin{figure*}
\begin{center}
\includegraphics[width=0.49\textwidth]{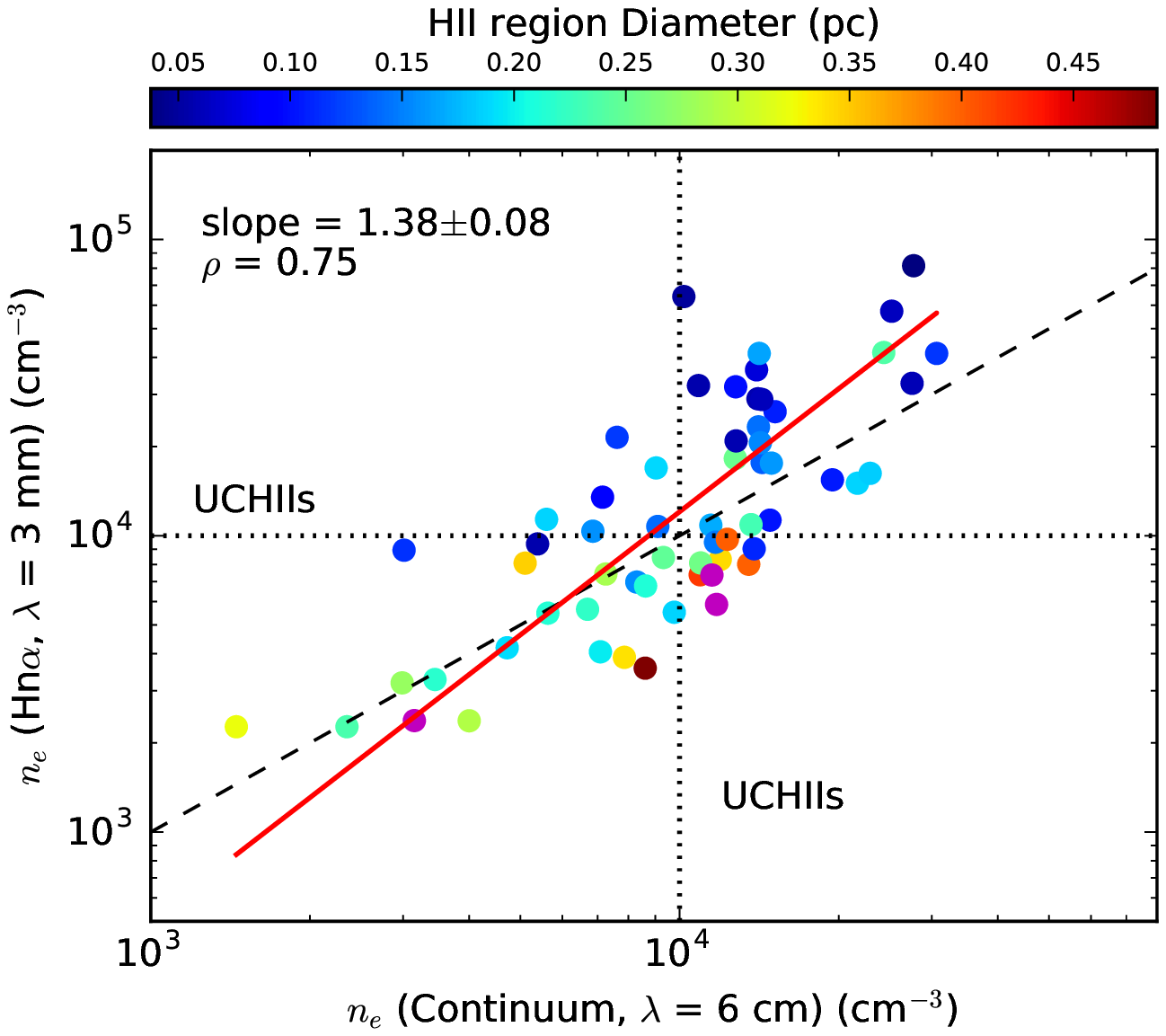}
\includegraphics[width=0.49\textwidth]{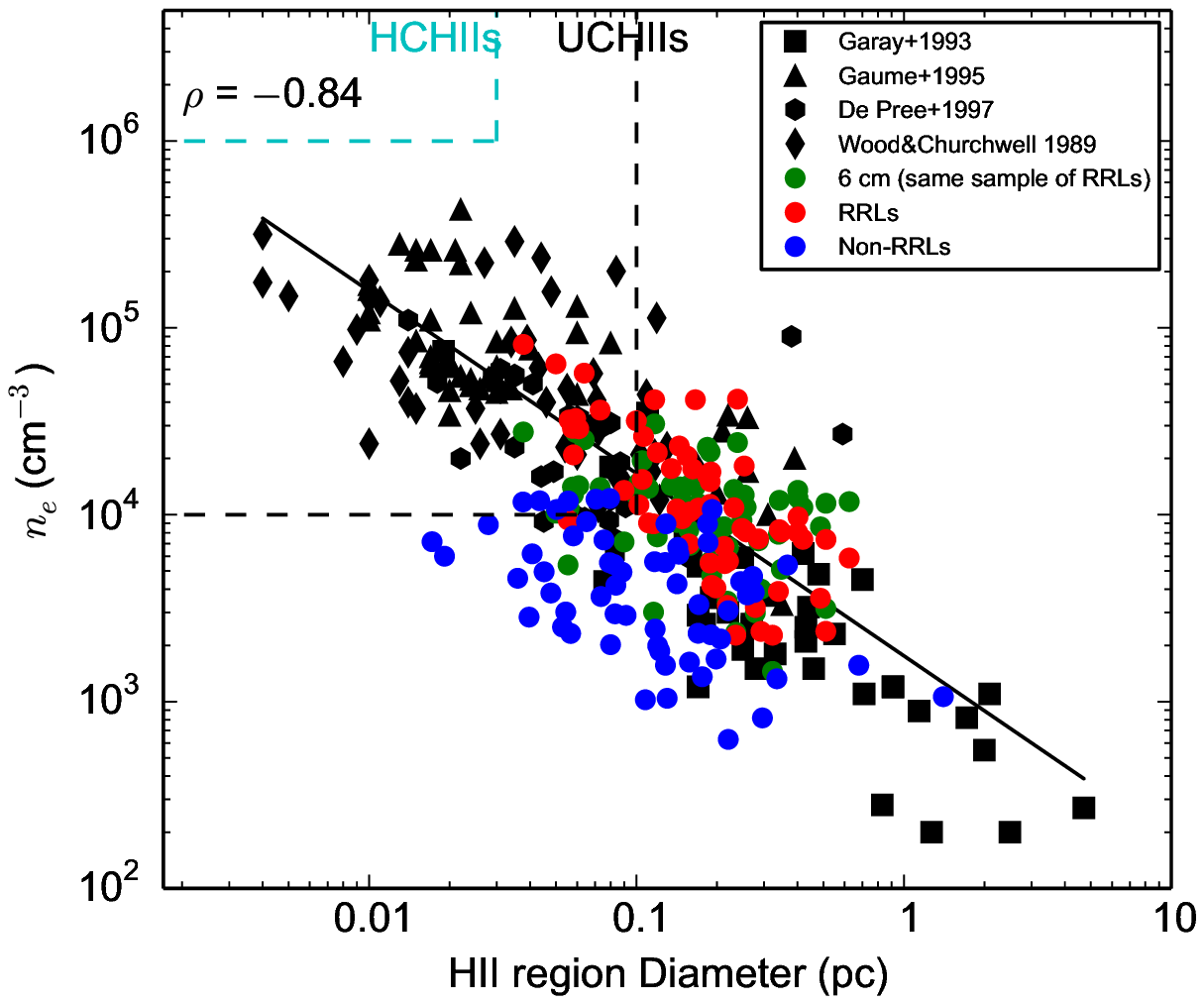}

\caption{\label{fig:ne} Left panel: $n_{e}$ derived from 6\,cm radio continuum emission versus $n_{e}$ from mm-RRLs. Equality is indicated by the black dashed line. The dotted lines show the thresholds of electron density for \uchii. The colors represent \hii\ region physical diameters in units of pc (see color bar) and the purple circles mark regions with diameter over 0.5\,pc. The red line represents the best-fit line determined by an xy-bisector fit to the data.\\
Right panel: $n_{e}$ as a function of \hii\ region diameter. The red and green symbols show the distribution of the same sample of mm-RRL detections but where the election densities are derived from the the mm-RRL emission and 6\,cm radio continuum emission, respectively. 
The black symbols indicate data from previous radio continuum surveys and the blue circles are radio sources toward which no mm-RRL emission is detected and where the electron density has been estimated from the of 6\,cm radio continuum emission. The black solid line represents the best-fit line determined by a least-squares polynomial fit with one degree to the data of the previous radio continuum surveys (black symbols) and this mm-RRL surveys (red symbols).}
\end{center}
\end{figure*}

\subsection{Emission measure and electron density}
\label{sec:em_ne}

In order to compare the physical parameters estimated by mm-RRL emission and 6\,cm radio continuum, we need to calculate the emission measure (EM) of the continuum sources. 
The synthesized beam brightness temperature is calculated by Eq. 2 in \citet{wood1989a}

\begin{equation}
\label{eq:tc}
T_{b} = \frac{S_{\nu} 10^{-29} c^{2}}{2 \nu^{2} k \Omega_{b}}
,\end{equation}

\noindent where $S_{\nu}$ is the peak flux density (mJy\,beam$^{-1}$), $\Omega_{b}$ is the solid angle of synthesized beam, $k_{\rm{b}}$ is the Boltzmann constant, 
$\nu$ is the rest frequency in Hz, and $c$ is the speed of light in m\,s$^{-1}$. The solid angle $\Omega_{b}$ is calculated from $\Omega_{b} = 1.133\theta_{b}^{2}\,(sr)$,   where $\theta_{b}$ is $1$\farcs$5$ for the CORNISH sources \citep{purcell2013}, $2$\farcs$5$ for the RMS sources \citep{urquhart2007}, and $4$\farcs$0$ for the sources identified by \cite{becker1994}. We assume $\tau_{C} \ll 1$ at 5\,GHz and the Rayleigh-Jeans approximation,
and thus $\tau_{C}$ = -ln$\left(1-T_{b}/T_{e}\right)$,

\begin{equation}
\label{eq:em_c}
{\rm EM}_{\rm cont}\,({\rm{cm}}^{-6}\,{\rm{pc}}) = \frac{\tau_{C}}{8.235\times10^{-2}a(\nu, T_{e})T_{e}^{-1.35}\nu^{-2.1}}
,\end{equation}

\noindent where $a$ is a correction factor of order unity \citep{mezger1967}, which is 0.9938 at an electron density ($T_{e}$) of 10,000\,K \citep{wood1989a} and $\nu$ = 5\,GHz. 
Finally, in order to estimate an electron density ($n_{e}$), we use the physical diameters of the 6\,cm radio continuum sources. If there are multiple continuum sources within the Mopra/IRAM beams, we estimate a new \hii\ region diameter from the combined area of radio continuum sources, i.e., $\Delta s$ = $\sqrt{\sum_{n=1}^{n}\Delta s_{n}^{2}}$. The $n_{e}$ is then given by

\begin{equation}
\label{eq:ne}
n_{e} = \sqrt{\frac{\rm EM}{\Delta s}}
.\end{equation}

Since we  confirmed that the linewidths of detected mm-RRLs are dominated by Doppler broadening (i.e., in LTE) (Sect.\,\ref{sec:lines}) and that the mm-RRLs seem to be optically thin (Sect.\,\ref{sec:radio_rrl_intensity}), we can determine the EM from the velocity-integrated intensity of the measured mm-RRL \citep{rohlfs2000}.  
We estimate the EM using the following equation that takes account of the beam dilution effect,

\begin{equation}
\label{eq:em_rrl}
{\rm EM}_{\rm RRL}\,({\rm{cm}}^{-6}\,{\rm{pc}}) = \frac{T_{e}^{3/2}}{576}\left(\frac{\theta_{\rm MB}^{2} + \theta_{s}^{2}}{\theta_{s}^{2}}\right) \left(\frac{\nu_{0}}{{\rm{GHz}}}\right) \left(\frac{\int T_{\rm MB}\,\Delta V}{K\,\kms}\right)
,\end{equation}

\noindent where the beam filling factor, $f_{beam}$, is given by $\theta_{s}^{2}/(\theta_{\rm MB}^{2}+\theta_{s}^{2})$ in which $\theta_{s}$ is the angular diameter of the radio continuum source and $\theta_{\rm MB}$ is the antenna main-beam size. 
The angular diameter also presents a summed size for  the physical diameters of radio continuum sources. 
Here we use the same $T_{e}$ as with the radio continuum data.\\

In the left panel of Fig.\,\ref{fig:ne} we present a comparison of the $n_{e}$ calculated independently from the radio continuum and mm-RRL emission. There is a strong correlation between these two measurements with a correlation coefficient of  $\rho=0.75$ with $p-$value $\ll 0.001$. The xy-bisector fit to these data gives a slope of 1.38\,$\pm$\,0.08. There is a noticeable trend for the more compact radio sources to have higher $n_{e}$ in the measurement by the mm-RRL emission, which is  expected since they are likely to be less optically thin.

In the right panel of Fig.\,\ref{fig:ne} we show the relationship between $n_{e}$ and the \hii\ region diameter. The black symbols indicate the distribution of \hii\ regions reported in the literature, while the red and blue symbols show the properties of mm-RRL loud and quiet \hii\ regions discussed in this paper. There is clearly a strong correlation between these parameters ($\rho=-0.84$ with $p-$value $\ll 0.001$).

As previously discussed, we have found that the detected mm-RRLs tend to be associated with brighter \hii\ regions, many of which are driven by  O-type stars.  Fig.\,\ref{fig:ne} reveals that the  mm-RRLs tend to be associated with \hii\ regions that are more evolved than the \uchii\ region stage. They also cover a similar range of physical scales to the mm-RRL quiet \hii\ regions, but have significantly higher $n_{e}$, which suggests that they are driven by more massive stars. Comparing these results to the typical $n_{e}$ and diameters of \hchii\ (cyan dashed line; diameter $\leq$ 0.03\,pc and $n_{e}$ $\geq$ 10$^{5}$ cm$^{-3}$) and \uchii\ (black dashed line; diameter $\leq$ 0.1\,pc and $n_{e}$ $\geq$ 10$^{4}$ cm$^{-3}$) \citep{kurtz2000}, we find that most of the mm-RRL sources correspond to compact \hii\ regions with a few located in the  \uchii\ region part of the parameter space. We note that none is located in the  \hchiis\ parameter space; however, this plot only includes mm-RRL sources that have been matched with 6\,cm radio continuum emission and so may be biased away from \hchii\ regions as these tend to be optically thick at this frequency and are therefore less likely to be detected. As discussed in Sect.\,\ref{sec:counterparts}, we have identified a number of new \hii\ region candidates, and although the nature of these objects needs to be confirmed, this sample does include a few good potential \hchii\ region candidates.

The results obtained from the mm-RRL parameters are consistent with the results determined from the radio continuum. This demonstrates the feasibility and complementarity of using mm-RRL observations to identify and parameterize compact \hii\ regions. Furthermore, mm-RRL observations have a couple of distinct advantages over conventional searches (e.g., radio continuum and mid-IR color selection) as all \hii\ regions should be optically thin at 3 mm and provide velocity information that can be used to derive distances and identify their natal molecular cloud.

\subsection{Potential young \hii\ regions}
\label{sec:hchii}

For a while, 
high-mass stars during their early evolutionary phase as HC \hii\ regions still show activities such as accretion, infall, and outflow. 
As discussed in Sect.\,\ref{sec:em_ne}, although we have found that the properties of most sources are indicative of compact and \uchii\ regions,  we have identified three cases in which broad non-thermal motions mark potential young \hii\ regions. 

In Sect.\,\ref{sec:lines}, we  report broad mm-RRLs from nine clumps that have linewidths broader than 40\,\kms. Roughly 30\% of \uchiis\ and 50\% of \hchiis\ are known to be associated with such broad RRLs observed at cm-wavelengths \citep{jaffe1999}.
Such broad RRL objects (BRLOs) are thought to be linked to a limited period having ionized outflows before the more evolved \uchii\ phase \citep{jaffe1999,sewilo2004b,sewilo2008,keto2008a, zhang2014,guzman2014}.
Interestingly, one of the nine  BRLOs, N73, has a velocity offset of its mm-RRL velocity with respect to its systemic velocity. This could be due to the influence of contamination by nearby \hii\ regions. Nevertheless, N73 is not only associated with a bright WISE 22\,\mum\ and unresolved radio continuum sources but is also  coincident with an ionized outflow candidate reported by \cite{guzman2012}. The ionized outflow candidate was explained to be an optically thick, expanding \hchii\ \citep{purser2016}. 
However, there is still no satisfactory explanation that is consistent with the origin of such BRLOs, although ionized outflows, disk winds, bow shocks, champagne flows, and inflows models \citep{jaffe1999} have been suggested to explain this phenomenon.

The second piece of  evidence for the extreme youth of some \hii\ regions is the existence of stimulated (maser) emission in the mm-RRLs of N49 and N55. The mm-RRLs of these two clumps seem to be weakly enhanced by maser amplification.
Radio recombination line maser emission is a very rare phenomenon and to date only two sources, WMC\,349A and Mon\,R2 \citep{martin-pintado1989,jimenez-serra2013}, have been  confirmed as RRL maser sources.
According to these studies, the RRL maser phenomenon is linked to the structure and kinematics of the internal ionized gas rather than the nature of the source, and therefore the maser emission could be expected toward some young \hii\ regions showing high internal electron density that are modified by ionized stellar winds. Observing higher frequency, i.e., lower $n$ (sub)mm-RRLs, would also be helpful; such observations toward MWC\,349A have revealed that the line shape becomes more and more asymmetric with decreasing $n$, developing a pronounced double-peaked shape \citep{thum1995,martin-pintado2002,jimenez-serra2013}.  
Interferometric observations are needed to confirm the RRL masers, similar to the confirmation of Mon R2 using high-resolution observations \citep{jimenez-serra2013}.  

We have found mm-RRLs toward eight clumps without a radio counterpart. In particular, some of these (e.g., N37 as shown in Fig.\,\ref{fig:hchii}) do not show extended 22\,\mum\ or 8\,\mum\ emission, but are associated with various masers such as OH, \water, Class I \& II \methanol. 
Source N37, also known as W43-MM1, is considered to be a hot core and has multiple outflows \citep{motte2003,sridharan2014}. 
In addition, wing features seen in SiO emission, like the case of N30, \citep{csengeri2016_sio} support the existence of outflows. Regardless of the absence of radio counterparts, the detections of the maser emission and the wings of SiO suggest that the detected mm-RRLs are produced from embedded young \hii\ regions in the clumps, but further mm- and submm-RRL observations at the high resolution and sensitivity provided by interferometers are necessary to study their properties in detail.

\section{Summary and conclusions}\label{sec:summary}

We  carried out 3 mm spectral line observations toward 976 dense clumps identified by the ATLASGAL survey ($-$60\degr\ $\leq \ell \leq +$60\degr). The sample was carefully selected to include a wide range of evolutionary stages from starless clumps to protostellar and compact \hii\ region stages. The observed spectra included a number of mm-RRLs and thermal molecular line transitions allowing a range of physical and environmental conditions to be investigated at the same angular resolution.

We have detected H$n\alpha$ mm-RRL emission toward 178 ATLASGAL clumps;  H$n\beta$,  H$n\gamma$, and H$n\delta$ transitions have been detected toward 65, 23, and 22 clumps, respectively. Inspecting mid-IR images and high-resolution radio continuum surveys, we are able to associate 134 of these mm-RRL detections with compact and \uchii\ regions previously identified in the literature. This represents the largest mm-RRL sample so far reported and provides a new method of identifying and parameterizing \hii\ regions. Comparing the radio and mid-IR fluxes of the mm-RRL quiet and loud sources, we find that they are associated with the brightest continuum sources. 
Thus, because of the limited sensitivity of our observations we were not able to detect mm-RRLs toward an additional  $\sim${126} \hii\ regions. Analysis of the ratios of the different mm-RRL transitions reveals that the lines are formed under LTE conditions.  

Comparing the systemic velocity of the molecular material and the velocity of the ionized gas traced by the mm-RRL emission ($\sigma <$\,5\,\kms) reveals a strong correlation between the two, which is consistent with the mm-RRL emitting \hii\ regions still being associated with their parental molecular clumps. We  expected feedback from the embedded \hii\ regions to have a direct effect on the global turbulent motion in their molecular clumps. Comparing the \hco\ linewidth of the mm-RRL associated and unassociated clumps, we find that the associated clumps tend to be more turbulent. However, we cannot reject the possibility that this is due to the Larson size-line width relation \citep{larson1981}.

We find a strong correlation between the integrated 6\,cm radio continuum and mm-RRL emissions (the correlation coefficient $\rho=0.70$). This result implies that the 6\,cm continuum and mm-RRL emissions are tracing the same ionized nebula. We also find that the Lyman continuum fluxes for these mm-RRL source are associated with evolved \hii\ regions driven by late O-type stars and the reason for large \hii\ region diameters ($\sim$0.5\,pc) is that the \hii\ regions driven by O-type stars expand faster than B-type stars.

Of the remaining 44 mm-RRL detections that are not confirmed with radio counterparts, 36 are thought to be associated with nearby evolved \hii\ regions, while 8 detections are considered to be potential new \hii\ region candidates; these will need to be confirmed by future high-resolution and high-sensitivity radio continuum observations. 

In this paper we have explored a new method of identifying compact \hii\ regions using mm-RRL emission. We have identified 142 genuine \hii\ regions (including the 8 potential \hii\ regions) and analysis of the line parameters has produced results that are consistent with those obtained from the radio continuum emission. We have therefore demonstrated that mm-RRLs are a viable and complementary method of identifying compact \hii\ regions and investigating their physical properties. Furthermore, mm-RRL observations have some inherent advantages over studies that merely analyze continuum emission in that they provide velocity information, which can be used to match the regions to  their host clumps and to determine their Galactic locations.

\begin{acknowledgements}
We would like to thank the referee for their constructive comments and suggestions that have helped to improve this paper.
This document was produced using the Overleaf web application, which can be found at www.overleaf.com. Won-Ju\, Kim was supported for this research through a stipend from the International Max Planck Research School (IMPRS) for Astronomy and Astrophysics at the Universities of Bonn and Cologne. 
Won-Ju\,Kim acknowledges partial support through the Bonn-Cologne Graduate School of Physics and Astronomy.
The ATLASGAL project is a collaboration between the Max-Planck-Gesellschaft, the European Southern Observatory (ESO), and the Universidad de Chile. It includes projects E-181.C-0885, E-078.F-9040(A), M-079.C-9501(A), M-081.C-9501(A), plus Chilean data.
\end{acknowledgements}

\begin{appendix}

\section{Comparison of cm- and mm-RRLs detected toward \hii\ regions}

So far, RRL studies with single-dish telescopes and interferometers have mostly  focused on RRLs at cm wavelength.
There are only a few mm-RRL surveys that  were conducted toward small samples biased toward a specific evolutionary phase such as \uchii\ or \hchii\ regions.
Our new survey   investigates for the first
time  the general properties of mm-RRLs  toward a large sample and in this Appendix, we  take a look at the overall properties of mm-RRLs in comparison with cm-RRLs of previous large surveys.

\begin{figure}
\centering
\includegraphics[width=0.5\textwidth]{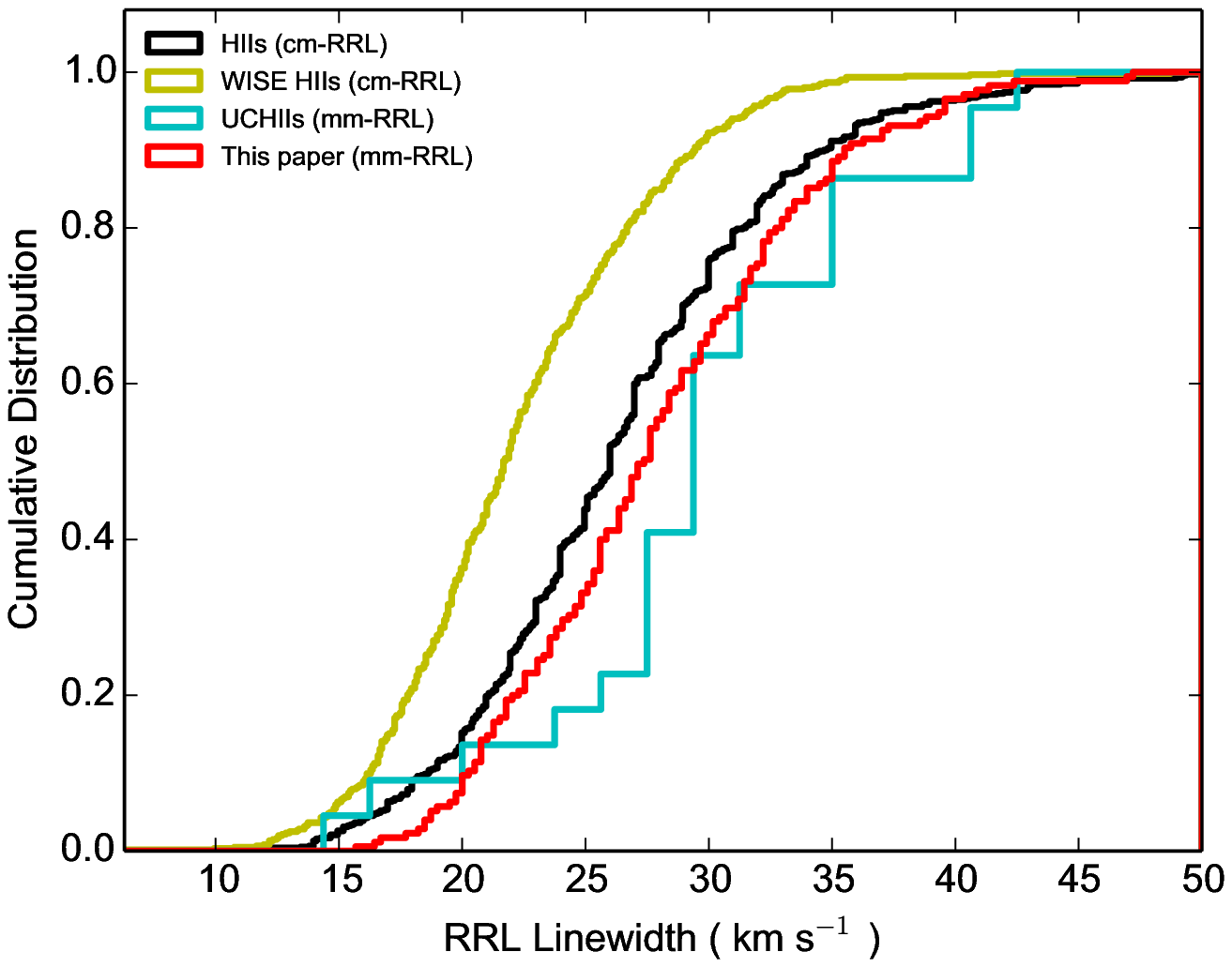}
\caption{\label{fig:rrl_survey} Cumulative distributions of RRL linewidths for this paper (red line) and previous surveys (yellow line: \citealt{anderson2011}; black line: \citealt{lockman1989,caswell1987}; and cyan line: \citealt{churchwell2010}).}
\end{figure}

Table\,\ref{tb:rrl_surveys} is a summary of cm- and mm-RRL surveys used in this comparison.
The RRLs surveys were carried out at different wavelengths with various beam sizes and targeted on different evolutionary phases of \hii\ regions, from \uchii\ to extended \hii\ regions. The 
\hii\ regions of \cite{caswell1987} and \cite{lockman1989} were classified by 6\,cm radio continuum surveys with single-dish telescopes, which have a poor spatial resolution for identifying  small \hii\ regions. These \hii\ regions could be in different evolutionary phases of \hii\ region. Therefore, in comparisons they are considered to be general \hii\ regions that are denoted by \hiis\ here. 
Fig.\,\ref{fig:rrl_survey} shows the cumulative distributions of RRL linewidths, in which the \hiis\ (a black distribution) combines cm-RRLs from \cite{lockman1989} and \cite{caswell1987}.

\begin{table}
\small
\centering
\begin{center}\caption{\label{tb:rrl_surveys} Hydrogen radio recombination line surveys.}
\begin{minipage}{1.2\linewidth}
\begin{tabular}{l c c c c }
\hline \hline
Survey & H$n\alpha$ transition & \multicolumn{1}{c}{$\lambda$} & Beam size & No. \\
\hline
This paper  & H39$\alpha$ $-$ H42$\alpha$ & 3\,mm & 29$''$ $-$ 36$''$  & 178 \\
Churchwell+10 & H30$\alpha$ & 1.3\,mm& 12$''$ & 25\\
Lockman+89      & H87$\alpha$ $-$ H88$\alpha$ & 3\,cm & 180$''$  & 462 \\
Anderson+11     & H88$\alpha$ $-$ H93$\alpha$ & 4\,cm & 73$''$ $-$ 89$''$ &  603 \\
Caswell+87      & H109$\alpha$ $-$ H110$\alpha$ & 6\,cm & 264$''$ & 317 \\
\hline
\end{tabular} 
\end{minipage}
\end{center}
\end{table}

\begin{figure}
\centering
\includegraphics[width=0.5\textwidth]{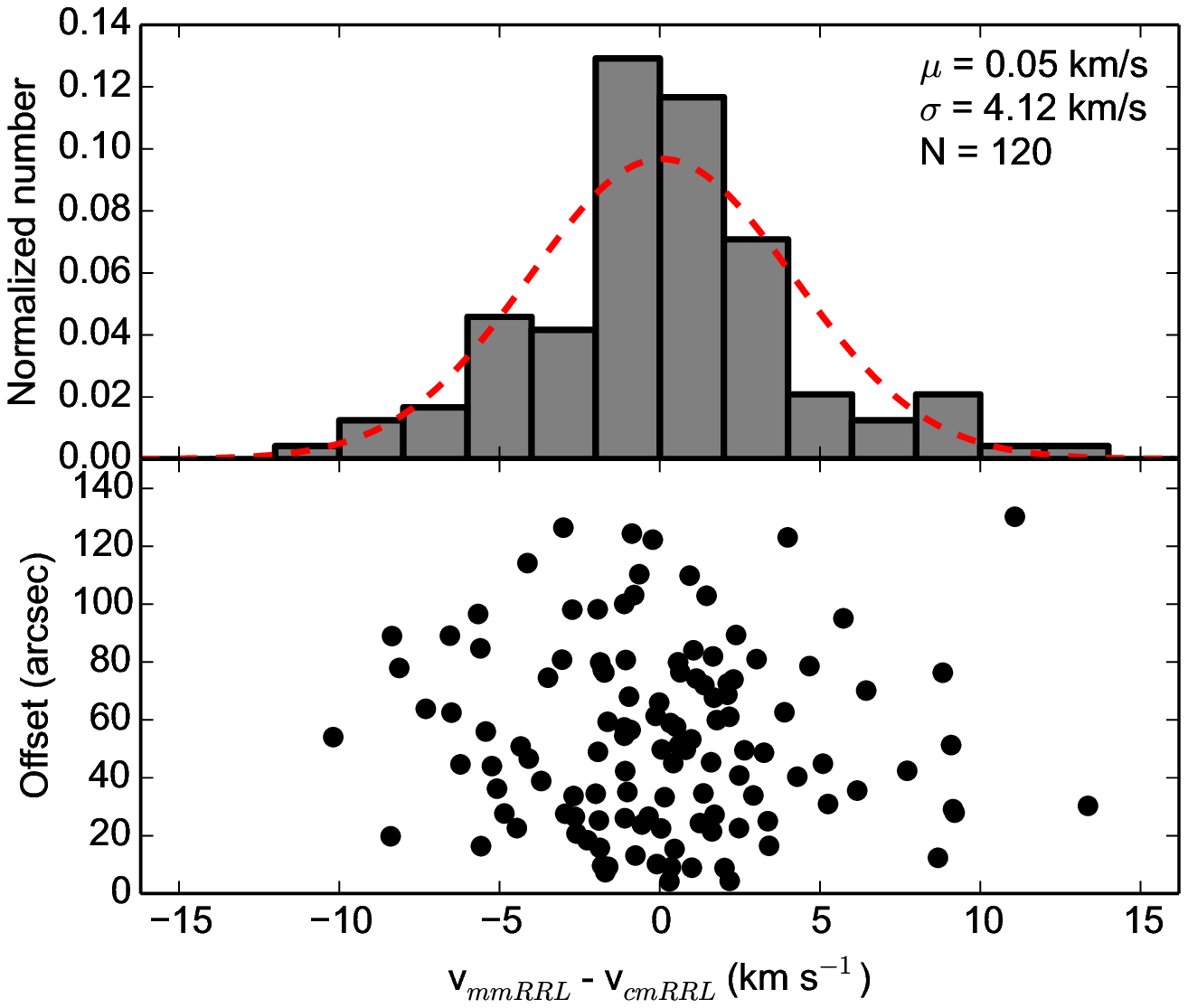}
\vskip -0.4cm
\includegraphics[width=0.5\textwidth]{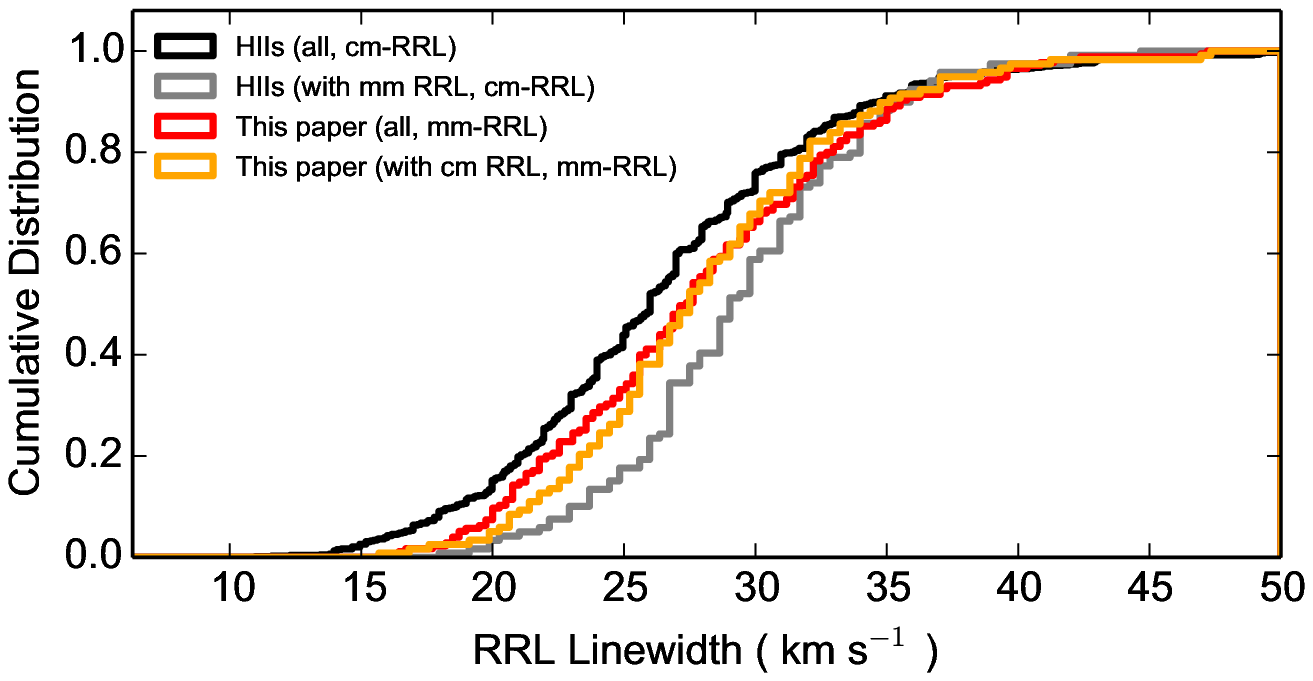}
\caption{\label{fig:mm_cm_rrl} Upper panel: Velocity difference between mm-RRL and cm-RRL for sources where the beams of the surveys overlap (top plot). The velocity difference as a function of an offset between the mm-RRL and cm-RRL pointing centers (bottom plot). Lower panel: Cumulative distributions of cm- (gray) and mm-RRLs (orange) for clumps where the velocity difference between the cm- and mm-RRLs is within 15\,\kms.}
\end{figure}

In Fig.\,\ref{fig:rrl_survey}, there is an obvious difference between linewidths of RRLs of the WISE selected \hiis\ (WISE \hiis) from \citet{anderson2011} and the other samples.   
The linewidths of WISE \hiis\ are significantly narrower (mean width of 22.3\,\kms) than the other surveys with mean values between 26.8 and 33.4\,\kms.
Kolmogorov-Smirnov tests are able to reject the null hypothesis, that the linewidths of the WISE \hii\ regions are similar to the other surveys, with very small $p-$values $\ll$ 0.001, implying differences in the properties of the \hii\ regions samples. 
In fact, the WISE \hiis, which are classified by 8\,\mum\ emission in GLIMPSE IRAC maps, include ionized bubbles, as well as more evolved \hii\ regions \citep{anderson2011}.
The sample (\uchiis) of  \cite{churchwell2010} seems to be offset from   the cm \hiis\ and from the sources discussed in this paper. 
However, the size of the sample is too small to be reliable and a null hypothesis that all of three samples are similar cannot be rejected because of a $p-$value (0.03) greater than 1\%.

In Fig.\,\ref{fig:rrl_survey}, the cm-RRLs of \hiis\ and the mm-RRLs from this paper show similar linewidth distributions, although the cm-RRLs can be significantly affected by pressure broadening.
The histogram in the upper panel of Fig.\,\ref{fig:mm_cm_rrl} shows velocity differences that are  less than 15\,\kms\ toward the majority of sources (120 clumps) and therefore suggest that the cm- and mm-RRLs are tracing on common \hii\ regions. 
In the bottom plot of the upper panel the scatter dots show the distribution of offsets between observed positions of cm- and mm-RRLs as a function of the velocity difference. The scatter is homogeneous and further supports the idea that the cm- and mm-RRLs are emitted from the same region.

The cumulative distributions in the lower panel of the Fig.\,\ref{fig:mm_cm_rrl} show RRL linewidths of \hiis\ and this paper toward common \hii\ region which match in observed positions and velocities. In addition, the distribution for all sources are also displayed.
The distribution of selected mm-RRLs from this paper (orange, with a mean linewidth of 28.5$\pm$0.5 \,\kms) does not change much compared to the total sample of mm-RRLs (red, mean linewidth of 28.4$\pm$0.5\,\kms).
However, the distribution of selected cm-RRLs from the \hiis\ sample (gray, mean linewidth of 29.6\,$\pm$\,0.4\,\kms) moves toward broader linewidths compared with the distribution of the whole \hiis\ sample (black, mean linewidth of 26.8\,$\pm$\,0.3\kms).
A KS test for the linewidths of the selected cm- and mm-RRLs samples also shows that a null hypothesis is invalid with a $p-$value $\ll$ 0.001.
This suggests that the selected cm-RRLs are likely suffering from pressure broadening.

However, the cm-RRL surveys were carried out with 2\,$\sim$\,6 times larger beams than ours. 
The broader cm-RRLs can also be affected by larger scale motions of the ionized gas in the larger beams. Our beams cover gas with a size of 1\,pc at typical distance of 5\,kpc. 
On the other hand, the beams of the cm-RRL surveys can capture gas with a size of about 4\,$\sim$\,6\,pc.
Therefore, the linewidths of the cm-RRLs could also be a   result of the blending of more than one \hii\ region leading to the observed broader linewidths.

\end{appendix}

\bibliographystyle{aa}
\bibliography{ms_wjkim}

\end{document}